\documentclass[preprint,aps,12pt,showpacs,nofootinbib,tightenlines]{revtex4}
\usepackage{amsmath}
\usepackage{amssymb}
\usepackage{epsfig}
\usepackage{graphicx}
\begin{document}
\preprint{NJNU-TH-06-23}
%%%%%%%%%%%%%%%%%%%%%%%%%%%%%%%%%%%%%%%%%%%%%

\newcommand{\beq}{\begin{eqnarray}}
\newcommand{\eeq}{\end{eqnarray}}
\newcommand{\non}{\nonumber\\ }

\newcommand{\acp}{{\cal A}_{CP}}
\newcommand{\etap}{\eta^{(\prime)} }
\newcommand{\etapr}{\eta^\prime }
\newcommand{\pb}{\phi_B}
\newcommand{\pp}{\phi_{\pi}}
\newcommand{\pe}{\phi_{\eta}}
\newcommand{\pepr}{\phi_{\etap}}
\newcommand{\ppp}{\phi_{\pi}^P}
\newcommand{\pep}{\phi_{\eta}^P}
\newcommand{\peprp}{\phi_{\etap}^P}
\newcommand{\ppt}{\phi_{\pi}^t}
\newcommand{\pet}{\phi_{\eta}^t}
\newcommand{\peprt}{\phi_{\etap}^t}
\newcommand{\fb}{f_B }
\newcommand{\fpi}{f_{\pi} }
\newcommand{\feta}{f_{\eta} }
\newcommand{\fetap}{f_{\etap} }
\newcommand{\rpi}{r_{\pi} }
\newcommand{\re}{r_{\eta} }
\newcommand{\rep}{r_{\etap} }
\newcommand{\mb}{m_B }
\newcommand{\mop}{m_{0\pi} }
\newcommand{\moe}{m_{0\eta} }
\newcommand{\moep}{m_{0\etap} }

\newcommand{\psl}{ p \hspace{-1.8truemm}/ }
\newcommand{\nsl}{ n \hspace{-2.2truemm}/ }
\newcommand{\vsl}{ v \hspace{-2.2truemm}/ }
\newcommand{\epsl}{\epsilon \hspace{-1.8truemm}/\,  }

\def \epjc{ Eur.Phys.J. C }
\def \jpg{  J. Phys. G }
\def \npb{  Nucl. Phys. B }
\def \plb{  Phys. Lett. B }
\def \pr{  Phys. Rep. }
\def \prd{  Phys. Rev. D }
\def \prl{  Phys. Rev. Lett.  }
\def \zpc{  Z. Phys. C  }
\def \jhep{ J. High Energy Phys.  }

%%%%%%%%%%%%%%%%%%%%%%%%%%%%%%%%%%%%%%%%%%%%%%%%%%%%
%%
\title{Branching Ratio and CP Asymmetry of $B^0 \to \eta^{(\prime)} \eta^{(\prime)}$
Decays in the Perturbative QCD Approach}
\author{Zhen-jun Xiao\footnote{Electronic address: xiaozhenjun@njnu.edu.cn},
Dong-qin Guo and Xin-fen Chen }
\affiliation{Department of
Physics and Institute of Theoretical Physics, Nanjing Normal
University, Nanjing, Jiangsu 210097, P.R.China}
\date{\today}
\begin{abstract}
We calculate the CP averaged branching ratios and CP-violating
asymmetries for $B^0 \to \eta \eta, \eta \eta^\prime$ and
$\eta^\prime \eta^\prime$ decays by employing the perturbative QCD
(pQCD) factorization approach. The pQCD predictions for the
CP-averaged branching ratios are
$Br(B^0 \to \eta \eta) \approx 0.67 \times 10^{-7}$,
$Br(B^0 \to \eta \eta^\prime) \approx 0.18 \times 10^{-7}$,
and $Br(B^0 \to \eta^{\prime} \eta^{\prime}) \approx 0.11 \times 10^{-7}$,
which are consistent with currently available experimental upper limits.
We also predict large CP-violating asymmetries for the considered three decay modes,
which can be tested by the future B meson experiments.
\end{abstract}

\pacs{13.25.Hw, 12.38.Bx, 14.40.Nd}

\maketitle

\section{Introduction}

The two-body charmless B meson decays provide a good place
for testing the standard model (SM), studying CP violation of B meson system,
exploring the rich quantum chromodynamics (QCD) of strong interaction
and searching for the signal or evidence of new physics beyond the SM \cite{cpv}.
Up to now, many $B \to M_1 M_2$ decays (where $M_i$ refers to the light pseudo-scalar
or vector mesons )
have been measured experimentally \cite{bexp1,hfag06} with good accuracy,
calculated and studied phenomenologically in the QCD factorization (QCDF) approach
\cite{bbns99,bn03b,du03,yy01}, the perturbative QCD (pQCD) factorization approach
\cite{luy01,kls01,li01,kklls04,li05,ekou1,ekou2,liu06,wang06},
or the soft collinear effective theory (SCET) \cite{wz0601}.

Among various  $B \to M_1 M_2$ decay channels, the decays involving the isosinglet
$\eta$ or $\etapr$ mesons in the final state  have been studied extensively during
the past decade because of the so-called $K\etapr $ puzzle or other special
features. At present, we still do not know how large is
the gluonic content of the $\eta^\prime$ meson, and how to calculate reliably the
gluonic contributions to the  decay modes involving $\etapr$ meson.
It is still difficult to explain the observed pattern of
the branching ratios for $B \to K^{(*)} \eta^{(\prime)}$ decays.

In pQCD factorization approach, the $B \to K \etap$, $\rho \etap$ and $\pi \etap$
decays have been studied
in Refs.\cite{ekou1,ekou2,liu06,wang06}. In this paper, we would like to calculate the
branching ratios and CP asymmetries for the three $B \to \eta \eta, \eta \etapr$ and $\etapr \etapr$
decays by employing the low energy effective Hamiltonian \cite{buras96} and
the pQCD approach \cite{cl97,li2003,lb80}.
Besides the usual factorizable contributions, we here are able to evaluate the non-factorizable
and the annihilation contributions to these decays.
On the experimental side, the CP-averaged branching ratios of $B \to \eta^{(')} \eta^{(')}$ decays
have been measured very recently \cite{babar} in units of $10^{-6}$ (upper limits at $90\%$ C.L.):
\beq
Br(B^0 \to \eta \eta)&=& 1.1 ^{+0.5}_{-0.4} \pm 0.1\ \  (< 1.8), \label{eq:exp1}\\
Br(B^0 \to \eta\eta^\prime)&=&0.2^{+0.7}_{-0.5} \pm 0.4\ \  (< 1.7), \label{eq:exp2}\\
Br(B^0 \to \eta^\prime \eta^\prime)&=& 1.0^{+0.8}_{-0.6} \pm 0.1 \ \ (< 2.4).
\label{eq:exp3}
\eeq
The accuracy of the data is still low, only the upper limits can be used to compare with the theoretical
predictions.

This paper is organized as follows. In Sec.~\ref{sec:f-work}, we
give a brief review for the pQCD factorization approach. In
Sec.~\ref{sec:p-c}, we calculate analytically the related Feynman
diagrams and present the various decay amplitudes for the studied
decay modes. In Sec.~\ref{sec:n-d}, we show the numerical results
for the branching ratios and CP asymmetries of $B \to
\eta^{(\prime)} \eta^{(\prime)}$  decays and compare them with the
measured values or the theoretical predictions in QCDF approach. The
summary and some discussions are included in the final section.

\section{Theoretical framework}\label{sec:f-work}

For the non-leptonic B decays, the dominant theoretical uncertainty comes from the
evaluation of the hadronic matrix element $<M_1 M_2|O_i|B>$.
Now there are two popular factorization approaches being used to calculate
the matrix elements: the QCDF approach \cite{bbns99}
and the pQCD approach \cite{cl97,li2003,lb80}.
The pQCD approach has been developed earlier from the QCD
hard-scattering approach \cite{lb80}. Some elements of this
approach are also present in the QCD factorization approach \cite{bbns99,bn03b}.
The two major differences between these two approaches are
(a) the form factors are calculable perturbatively
in pQCD approach, but taken as the input parameters extracted from
other experimental measurements in the QCDF approach; and (b) the
annihilation contributions are calculable and play an important
role in producing CP violation for the considered decay modes in
pQCD approach, but it could not be evaluated reliably in QCDF
approach. Of course, one should remember that  the assumptions
behind the pQCD approach, specifically the possibility to
calculate the form factors perturbatively, are still under
discussion~\cite{ds02}. More efforts are needed to clarify these problems.

In  pQCD approach, the decay amplitude is separated into soft, hard, and
harder dynamics characterized by different energy scales $(t, m_b,M_W)$.
It is conceptually written as the convolution,
\beq {\cal
A}(B \to M_1 M_2)\sim \int\!\! d^4k_1 d^4k_2 d^4k_3\ \mathrm{Tr}
\left [ C(t) \Phi_B(k_1) \Phi_{M_1}(k_2) \Phi_{M_2}(k_3)
H(k_1,k_2,k_3, t) \right ],
\label{eq:con1}
\eeq
where $k_i$'s are momenta of light quarks included in each mesons,
and $\mathrm{Tr}$ denotes the trace over Dirac and color indices.
$C(t)$ is the Wilson coefficient which results from the radiative
corrections at short distance. In the above convolution, $C(t)$
includes the harder dynamics at larger scale than $M_B$ scale and
describes the evolution of local $4$-Fermi operators from $m_W$
(the $W$ boson mass) down to $t\sim\mathcal{O}(\sqrt{\bar{\Lambda}
M_B})$ scale, where $\bar{\Lambda}\equiv M_B -m_b$. The function
$H(k_1,k_2,k_3,t)$ is the hard part and can be  calculated perturbatively.
The function $\Phi_M$ is the wave function which describes hadronization of the quark and
anti-quark to the meson $M$. While the function $H$ depends on the
process considered, the wave function $\Phi_M$ is  independent of the specific
process. Using the wave functions determined from other well measured
processes, one can make quantitative predictions here.

Since the b quark is rather heavy we consider the $B$ meson at rest
for simplicity. It is convenient to use light-cone coordinate $(p^+,
p^-, {\bf p}_T)$ to describe the meson's momenta:
$p^\pm =(p^0 \pm p^3)/\sqrt{2}$ and ${\bf p}_T = (p^1, p^2)$.

Using the light-cone coordinates the $B$ meson and the two final
state meson momenta can be written as
\beq P_1 =\frac{M_B}{\sqrt{2}} (1,1,{\bf 0}_T), \quad
     P_2 =\frac{M_B}{\sqrt{2}} (1,0,{\bf 0}_T), \quad
     P_3 =\frac{M_B}{\sqrt{2}} (0,1,{\bf 0}_T),
\eeq
respectively, here the light meson masses have been neglected. Putting the light (anti-)
quark momenta in $B$, $\eta^\prime$ and $\eta$ mesons as $k_1$,
$k_2$, and $k_3$, respectively, we can choose
\beq
k_1 = (x_1 P_1^+,0,{\bf k}_{1T}), \quad k_2 = (x_2 P_2^+,0,{\bf k}_{2T}), \quad
k_3 = (0, x_3 P_3^-,{\bf k}_{3T}).
\eeq
Then, for $B \to \eta \etapr$ decay for example,  the integration over
$k_1^-$, $k_2^-$, and $k_3^+$ in eq.(\ref{eq:con1}) will lead to
\beq
{\cal A}(B \to \eta \eta^\prime) &\sim &\int\!\! d x_1 d x_2 d
x_3 b_1 d b_1 b_2 d b_2 b_3 d b_3 \non
&& \cdot \mathrm{Tr} \left [ C(t) \Phi_B(x_1,b_1) \Phi_(\eta^\prime)(x_2,b_2) \Phi_{\eta}(x_3,
b_3) H(x_i, b_i, t) S_t(x_i)\, e^{-S(t)} \right ], \label{eq:a2}
\eeq
where $b_i$ is the conjugate space coordinate of $k_{iT}$, and
$t$ is the largest energy scale in function $H(x_i,b_i,t)$. The
large logarithms $\ln (m_W/t)$ are included in the Wilson coefficients
$C(t)$. The large double logarithms ($\ln^2 x_i$) on the
longitudinal direction are summed by the threshold resummation ~\cite{li02},
and they lead to $S_t(x_i)$ which smears the end-point
singularities on $x_i$. The last term, $e^{-S(t)}$, is the Sudakov
form factor which suppresses the soft dynamics effectively
~\cite{soft}. Thus it makes the perturbative calculation of the hard
part $H$ applicable at intermediate scale, i.e., $M_B$ scale. We
will calculate analytically the function $H(x_i,b_i,t)$ for the considered
decays in the first order in $\alpha_s$ expansion
and give the convoluted amplitudes in next section.

\subsection{ Wilson Coefficients}\label{ssec:w-c}

For the two-body charmless B meson decays, the related weak effective
Hamiltonian $H_{eff}$ can be written as \cite{buras96}
\beq
\label{eq:heff} {\cal H}_{eff} = \frac{G_{F}} {\sqrt{2}} \, \left[
V_{ub} V_{ud}^* \left (C_1(\mu) O_1^u(\mu) + C_2(\mu) O_2^u(\mu)
\right) - V_{tb} V_{td}^* \, \sum_{i=3}^{10} C_{i}(\mu) \,O_i(\mu)
\right] \; ,
\eeq
where $C_i(\mu)$ are Wilson coefficients at the renormalization scale $\mu$ and $O_i$
are the four-fermion operators for the case of $b \to d $ transition,
\beq
\begin{array}{llllll}
O_1^{u} & = &  \bar d_\alpha\gamma^\mu L u_\beta\cdot \bar u_\beta\gamma_\mu L b_\alpha\ ,
&O_2^{u} & = &\bar d_\alpha\gamma^\mu L u_\alpha\cdot \bar
u_\beta\gamma_\mu L b_\beta\ , \\
O_3 & = & \bar d_\alpha\gamma^\mu L b_\alpha\cdot \sum_{q'}\bar
 q_\beta'\gamma_\mu L q_\beta'\ ,   &
O_4 & = & \bar d_\alpha\gamma^\mu L b_\beta\cdot \sum_{q'}\bar
q_\beta'\gamma_\mu L q_\alpha'\ , \\
O_5 & = & \bar d_\alpha\gamma^\mu L b_\alpha\cdot \sum_{q'}\bar
q_\beta'\gamma_\mu R q_\beta'\ ,   & O_6 & = & \bar
d_\alpha\gamma^\mu L b_\beta\cdot \sum_{q'}\bar
q_\beta'\gamma_\mu R q_\alpha'\ , \\
O_7 & = & \frac{3}{2}\bar d_\alpha\gamma^\mu L b_\alpha\cdot
\sum_{q'}e_{q'}\bar q_\beta'\gamma_\mu R q_\beta'\ ,   & O_8 & = &
\frac{3}{2}\bar d_\alpha\gamma^\mu L b_\beta\cdot
\sum_{q'}e_{q'}\bar q_\beta'\gamma_\mu R q_\alpha'\ , \\
O_9 & = & \frac{3}{2}\bar d_\alpha\gamma^\mu L b_\alpha\cdot
\sum_{q'}e_{q'}\bar q_\beta'\gamma_\mu L q_\beta'\ ,   & O_{10} &
= & \frac{3}{2}\bar d_\alpha\gamma^\mu L b_\beta\cdot
\sum_{q'}e_{q'}\bar q_\beta'\gamma_\mu L q_\alpha'\ ,
\label{eq:operators}
\end{array}
\eeq
where $\alpha$ and $\beta$ are the $SU(3)$ color indices; $L$
and $R$ are the left- and right-handed projection operators with
$L=(1 - \gamma_5)$, $R= (1 + \gamma_5)$. The sum over $q'$ runs over
the quark fields that are active at the scale $\mu=O(m_b)$, i.e.,
$q'\epsilon\{u,d,s,c,b\}$. For the Wilson coefficients $C_i(\mu)$
($i=1,\ldots,10$), we will use the leading order (LO)
expressions, although the next-to-leading order (NLO)  results
already exist in the literature ~\cite{buras96}. This is the
consistent way to cancel the explicit $\mu$ dependence in the
theoretical formulae. For the renormalization group evolution of the Wilson coefficients from
higher scale to lower scale, we use the formulae as given in Ref.\cite{luy01}
directly.

\subsection{Wave Functions}\label{ssec:w-f}

In the resummation procedures, the $B$ meson is treated as a
heavy-light system. In general, the B meson light-cone matrix
element can be decomposed as ~\cite{grozin,bbns99}
\beq
&&\int_0^1\frac{d^4z}{(2\pi)^4}e^{i\bf{k_1}\cdot z}
   \langle 0|\bar{b}_\alpha(0)d_\beta(z)|B(p_B)\rangle \nonumber\\
&=&-\frac{i}{\sqrt{2N_c}}\left\{(\psl_B+m_B)\gamma_5
\left[\phi_B ({\bf k_1})-\frac{\nsl-\vsl}{\sqrt{2}}
\bar{\phi}_B({\bf k_1})\right]\right\}_{\beta\alpha}, \label{aa1}
\eeq
 where $n=(1,0,{\bf 0_T})$, and $v=(0,1,{\bf 0_T})$ are the
unit vectors pointing to the plus and minus directions,
respectively. From the above equation, one can see that there are
two Lorentz structures in the B meson distribution amplitude (DA).
They obey to the following normalization conditions
 \beq
 \int\frac{d^4 k_1}{(2\pi)^4}\phi_B({\bf k_1})
 =\frac{f_B}{2\sqrt{2N_c}}, ~~~\int \frac{d^4
k_1}{(2\pi)^4}\bar{\phi}_B({\bf k_1})=0.
 \eeq

In general, one should consider these two Lorentz structures in
calculations of $B$ meson decays. However, it can be argued that the
contribution of $\bar{\phi}_B$ is numerically small ~
\cite{luyang,kurimoto}, thus its contribution can be numerically
neglected. Therefore, we only consider the contribution of Lorentz
structure
\beq
\Phi_B= \frac{1}{\sqrt{2N_c}} (\psl_B +m_B) \gamma_5
\phi_B ({\bf k_1}), \label{bmeson}
\eeq
in our calculation. We use the same wave functions as in
Refs.~\cite{luy01,kls01,kurimoto}. Through out this paper, we
use the light-cone coordinates to write the four momentum as
($k_1^+,k_1^-, k_1^\perp$). In the next section, we will see that
the hard part is always independent of one of the $k_1^+$ and/or
$k_1^-$, if we make some approximations. The B meson wave function
is then the function of  variable $k_1^-$ (or $k_1^+$) and
$k_1^\perp$.
\beq
\phi_B (k_1^-, k_1^\perp)=\int d k_1^+ \phi
(k_1^+, k_1^-, k_1^\perp). \label{int}
\eeq

The wave function for $d\bar{d}$ components of $\eta^{(\prime)}$ meson are given as \cite{ekou2}
\beq
\Phi_{\eta_{d\bar{d}}}(P,x,\zeta)\equiv
\frac{i \gamma_5}{\sqrt{2N_c}} \left [ \psl \phi_{\eta_{d\bar{d}}}^{A}(x)+m_0^{\eta_{d\bar{d}}}
\phi_{\eta_{d\bar{d}}}^{P}(x)+\zeta m_0^{\eta_{d\bar{d}}} (
\vsl \nsl - v\cdot n)\phi_{\eta_{d\bar{d}}}^{T}(x) \right ],
\label{eq:ddbar}
\eeq
where $P$ and $x$ are the momentum and the momentum fraction of
$\eta_{d\bar{d}}$ respectively, while $\phi_{\eta_{d\bar{d}}}^A$,
$\phi_{\eta_{d\bar{d}}}^P$ and $\phi_{\eta_{d\bar{d}}}^T$ represent
the axial vector, pseudoscalar and tensor components of the wave
function respectively.  Following Ref.~\cite{ekou2}, we here also assume that the wave
function of $\eta_{d\bar{d}}$ is same as the $\pi$ wave function based on SU(3) flavor symmetry.
The parameter $\zeta$ is either $+1$ or $-1$ depending on the
assignment of the momentum fraction $x$.

The transverse momentum $k^\perp$ is usually converted to the $b$ parameter
by Fourier transformation.  The initial conditions of the function $\phi_i(x)$ with $i=(B,\eta, \eta')$
are of non-perturbative origin, satisfying the
normalization
\beq
\int_0^1\phi_i(x,b=0)dx=\frac{1}{2\sqrt{6}}{f_i}\;, \label{no}
\eeq
with $f_i$ the meson decay constants.

\section{Perturbative Calculations}\label{sec:p-c}

In this section, we will calculate and show the decay amplitude for each diagram including
wave functions. The hard part $H(t)$ involves the four quark operators and the necessary hard
gluon connecting the four quark operator and the spectator quark.  We first consider
$B \to \eta \etap $ decay mode as an example, and then extend our study to
$B \to \eta^\prime \eta^\prime$ decay.

\subsection{Decay amplitudes}

Similar to the $B \to \pi \etap$ decays in \cite{wang06} , there are 8 type diagrams
contributing to the  $B \to \eta \etap $ decays, as
illustrated in Figure 1. We first calculate the usual factorizable
diagrams (a) and (b). Operators $O_1$, $O_2$, $O_3$, $O_4$, $O_9$,
and $O_{10}$ are $(V-A)(V-A)$ currents, the sum of their amplitudes
is given as
\beq
F_{e\eta}&=& 8\pi C_F m_B^4\int_0^1 d x_{1} dx_{3}\,
\int_{0}^{\infty} b_1 db_1 b_3 db_3\, \phi_B(x_1,b_1) \non & &
\times \left\{ \left[(1+x_3) \pe(x_3, b_3) +(1-2x_3) \re
(\phi_\eta^p(x_3,b_3) +\phi_\eta^t(x_3,b_3))\right] \right. \non &&
\left.\quad  \cdot \alpha_s(t_e^1)\,
h_e(x_1,x_3,b_1,b_3)\exp[-S_{ab}(t_e^1)] \right.\non && \left.
+2\re \phi_\eta^p (x_3, b_3)
\alpha_s(t_e^2)h_e(x_3,x_1,b_3,b_1)\exp[-S_{ab}(t_e^2)] \right\}.
\label{eq:ab}
\eeq
where $\re=m_0^\eta/m_B$; $C_F=4/3$ is a color factor.
The function $h_e$, the scales $t_e^i$ and the Sudakov
factors $S_{ab}$ are displayed in Appendix \ref{sec:app1}.

\begin{figure}[t,b]
\vspace{-2 cm} \centerline{\epsfxsize=21 cm \epsffile{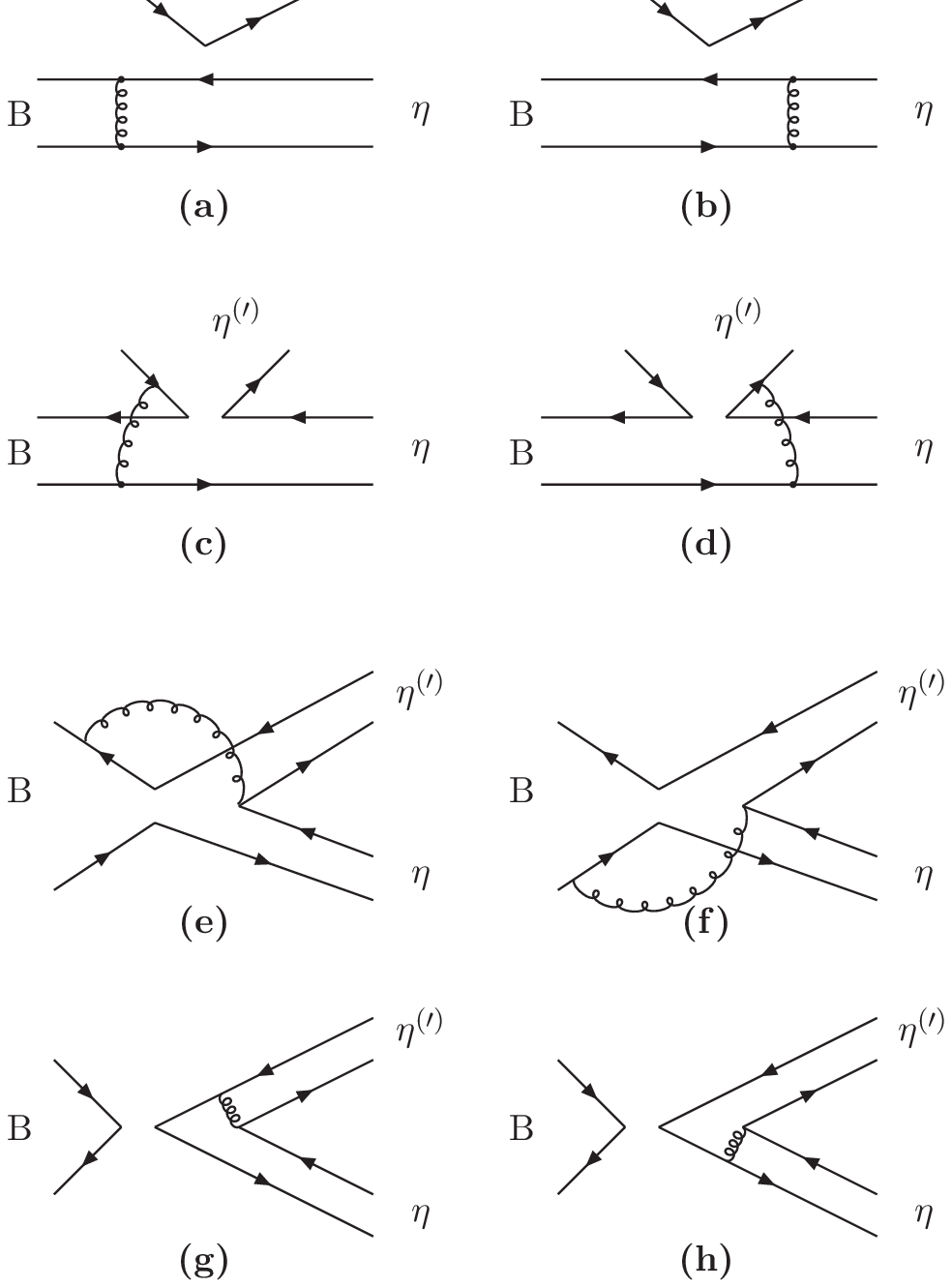}}
\vspace{-14cm} \caption{ Typical Feynman diagrams contributing to the $B\to
\eta\eta^{\prime}$  decays, where diagram (a) and (b) contribute to the $B\to \eta$ form
 factor $F_{0,1}^{B\to \eta}$.}
 \label{fig:fig1}
\end{figure}

The form factors of $B$ to $\eta$ decay, $F_{0,1}^{B\to \eta}(0)$,
can thus be extracted from the expression in Eq.~(\ref{eq:ab}), that
is
\beq
F_0^{B\to\eta}(q^2=0)=F_1^{B\to\eta}(q^2=0)= F_{e\eta}/m_B^2, \label{eq:f01}
\eeq
which is identical with that defined in Ref.~\cite{luyang}.

The operators $O_5$, $O_6$, $O_7$, and $O_8$ have a structure of
$(V-A)(V+A)$. In some decay channels, some of these operators
contribute to the decay amplitude in a factorizable way. Since only
the axial-vector part of $(V+A)$ current contribute to the
pseudo-scaler meson production, $ \langle \eta |V-A|B\rangle \langle
\eta^{\prime}|V+A | 0 \rangle = -\langle \eta |V-A |B  \rangle
\langle \eta^{\prime} |V-A|0 \rangle,$ that is
 \beq
 F_{e\eta}^{P1}=-F_{e\eta} \; .
 \eeq
For other cases, we need to do Fierz transformation  for the corresponding
operators to get right flavor and color structure for factorization to work. We may
get $(S-P)(S+P)$ operators from $(V-A)(V+A)$ ones. For
these $(S-P)(S+P)$ operators, Fig.~1(a) and 1(b) give
\beq
 F_{e\eta}^{P2}&=& 16\pi C_F m_B^4\rep \; \int_{0}^{1}d x_{1}d
x_{3}\,\int_{0}^{\infty} b_1d b_1 b_3d b_3\, \pb(x_1,b_1) \non & &
\times
 \left\{ \left[ \pe(x_3, b_3)+ \re((2+x_3) \pep (x_3, b_3)-x_3\pet(x_3, b_3))\right]
\right.  \non
& &\left. \cdot \alpha_s (t_e^1)  h_e
(x_1,x_3,b_1,b_3)\exp[-S_{ab}(t_e^1)]\right.  \non & &\left. \
   +\left[x_1\pe(x_3, b_3)-2(x_1-1)\re\pep (x_3, b_3)\right]
\right.  \non
& &\left. \cdot \alpha_s (t_e^2)
 h_e(x_3,x_1,b_3,b_1)\exp[-S_{ab}(t_e^2)] \right\} \; .
 \eeq

For the non-factorizable diagrams 1(c) and 1(d), all three meson
wave functions are involved. The integration of $b_3$ can be
performed using $\delta$ function $\delta(b_3-b_1)$, leaving only
integration of $b_1$ and $b_2$. For the $(V-A)(V-A)$ operators, the
result is
\beq
 M_{e\eta}&=& \frac{16 \sqrt{6}}{3}\pi C_F m_B^4
\int_{0}^{1}d x_{1}d x_{2}\,d x_{3}\,\int_{0}^{\infty} b_1d b_1 b_2d
b_2\, \pb(x_1,b_1) \pepr(x_2,b_2) \non
 & &\times
\left \{\left[2x_3\re \pet(x_3,b_1)-x_3 \phi_\eta(x_3,b_1)\right
]\right.\non
 & & \cdot \alpha_s(t_f) h_f(x_1,x_2,x_3,b_1,b_2)\exp[-S_{cd}(t_f)] \} \; .
\eeq

For the $(V-A)(V+A)$ operators the formulae are different. Here we
have two kinds of contributions from $(V-A)(V+A)$ operators:
$M_{e\eta}^{P1}$ and $M_{e\eta}^{P2}$ is for the  $(V-A)(V+A)$
and  $(S-P)(S+P)$ type operators respectively:
\beq
M_{e\eta}^{P1}&=&0 , \quad M_{e\eta}^{P2} = -M_{e\eta}\; .
\eeq

The factorizable annihilation diagrams 1(g) and 1(h) involve only
$\eta$ and $\etap$ wave functions. There are also three kinds of
decay amplitudes for these two diagrams. $F_{a\eta}$ is for
$(V-A)(V-A)$ type operators, $F_{a\eta}^{P1}$ is for $(V-A)(V+A)$
type operators, while $F_{a\eta}^{P2}$ is for $(S-P)(S+P)$ type
operators.
\beq
 F_{a\eta}^{P1}= F_{a\eta}&=&-8\pi C_F m_B^4\int_{0}^{1}dx_{2}\,d x_{3}\,
\int_{0}^{\infty} b_2d b_2b_3d b_3 \, \left\{ \left[x_3 \pe(x_3,b_3)
\pepr(x_2,b_2)\right.\right.\non &&\left.\left.+2 \re
\rep((x_3+1)\pep(x_3,b_3)+(x_3-1) \pet(x_3,b_3))
\peprp(x_2,b_2)\right] \right. \non && \left. \quad \cdot
\alpha_s(t_e^3) h_a(x_2,x_3,b_2,b_3)\exp[-S_{gh}(t_e^3)] \right.
\non && \left. -\left[ x_2 \pe(x_3,b_3) \pepr(x_2,b_2)
\right.\right.\non && \left. \left. \quad +2 \re \rep
((x_2+1)\peprp(x_2,b_2)+(x_2-1) \peprt(x_2,b_2)
)\pep(x_3,b_3)\right] \right. \non &&\left. \quad \cdot
\alpha_s(t_e^4)
 h_a(x_3,x_2,b_3,b_2)\exp[-S_{gh}(t_e^4)]\right \}\;
\eeq

\beq F_{a\eta}^{P2}&=& -16 \pi C_F m_B^4 \int_{0}^{1}d x_{2}\,d
x_{3}\,\int_{0}^{\infty} b_2d b_2b_3d b_3 \,\non &&\times \left\{
\left[x_3 \re (\pep(x_3, b_3)-\pet(x_3, b_3))\pepr(x_2,b_2)+2\rep
\pe(x_3,b_3) \peprp(x_2,b_2) \right]\right.
 \non
&&\left.\quad \cdot \alpha_s(t_e^3)
h_a(x_2,x_3,b_2,b_3)\exp[-S_{gh}(t_e^3)]\right.
 \non
 &&\left.+\left[x_2\rep(\peprp(x_2,b_2)-\peprt(x_2,b_2))\pe(x_3,b_3)+2\re
\pepr(x_2,b_2)\pep(x_3,b_3)\right] \right.\non &&\left.\quad \cdot
 \alpha_s(t_e^4)
 h_a(x_3,x_2,b_3,b_2)\exp[-S_{gh}(t_e^4)]\right\}\;.
 \eeq

For the non-factorizable annihilation diagrams 1(e) and 1(f), again
all three wave functions are involved. Here we have three kinds of
contributions. $M_{a\eta}$, $M_{a\eta}^{P1}$ and $M_{a\eta}^{P2}$
describe the contributions from $(V-A)(V-A)$, $(V-A)(V+A)$ and
$(S-P)(S+P)$ type operators respectively:
\beq
M_{a\eta}&=& \frac{16\sqrt{6}}{3}\pi C_F m_B^4\int_{0}^{1}d x_{1}d
x_{2}\,d x_{3}\,\int_{0}^{\infty} b_1d b_1 b_2d b_2\,
\phi_B(x_1,b_1)\non && \times \left \{ -\left \{x_2 \pe(x_3, b_2)
\pepr(x_2,b_2) \right.\right.\non
 & & \left.\left.
 +\re \rep \left [ (x_2+x_3+2) \peprp(x_2,b_2)
 +(x_2-x_3)\peprt(x_2,b_2) \right ] \pep(x_3,b_2)
\right.\right.\non
 & & \left.\left.
  +\re \rep \left [ (x_2-x_3)\peprp(x_3,b_2) +(x_2+x_3-2)\peprt(x_2,b_2)\right
  ]\pet(x_3,b_2)
\right\} \right. \non && \left. \quad \cdot
\alpha_s(t_f^3)h_f^3(x_1,x_2,x_3,b_1,b_2)\exp[-S_{ef}(t_f^3)]
\right.  \non && \left.
  + \left \{ x_3\pe(x_3,b_2) \pepr(x_2,b_2) \right.\right. \non
 && \left. \left.
 +\re \rep
 \left [ (x_2+x_3)\peprp(x_2,b_2)+(x_3-x_2)\peprt(x_2,b_2) \right ]\pep(x_3,b_2)
\right. \right. \non && \left.\left.
 +\re \rep\left [(x_3-x_2)\peprp(x_2,b_2)+(x_2+x_3)\peprt(x_2,b_2) \right ]\pet(x_3,b_2)
  \right \}
\right. \non && \left.
 \quad \cdot  \alpha_s(t_f^4)h_f^4(x_1,x_2,x_3,b_1,b_2)\exp[-S_{ef}(t_f^4) ]
 \right \}\; ,
 \eeq
 \beq
M_{a\eta}^{P1}&=& \frac{16\sqrt{6}}{3}\pi C_F m_B^4\int_{0}^{1}d
x_{1}d x_{2}\,d x_{3}\,\int_{0}^{\infty} b_1d b_1 b_2d b_2\,
\phi_B(x_1,b_1)
 \non
& &\times \left \{ \left[(x_3-2)
\re\pepr(x_2,b_2)(\pep(x_3,b_2)+\pet(x_3,b_2))-(x_2-2)\rep\pe(x_3,b_2)
\right.\right. \non && \left.\left.
(\peprp(x_2,b_2)+\peprt(x_2,b_2)) \right] \cdot
\alpha_s(t_f^3)h_f^3(x_1,x_2,x_3,b_1,b_2)\exp[-S_{ef}(t_f^3)]
\right. \non && \left.
 -\left[x_3\re\pepr(x_2,b_2)(\pep(x_3, b_2)+\pet(x_3,
b_2))\right. \right.\non && \left. \left.
 -x_2\rep\pe(x_3,b_2)(\peprp(x_2,b_2)+\peprt(x_2,b_2))\right]
\right. \non &&
\left.  \cdot
\alpha_s(t_f^4)h_f^4(x_1,x_2,x_3,b_1,b_2)\exp[-S_{ef}(t_f^4)] \right \}
 \; ,
\eeq
\beq
 M_{a\eta}^{P2}&=& \frac{16\sqrt{6}}{3}\pi C_F m_B^4\int_{0}^{1}d x_{1}d
x_{2}\,d x_{3}\,\int_{0}^{\infty} b_1d b_1 b_2d b_2\,
\phi_B(x_1,b_1)\non && \times \left \{ \left \{x_3 \pe(x_3, b_2)
\pepr(x_2,b_2) \right.\right.\non
 & & \left.\left.
 +\re \rep \left [ (x_2+x_3+2) \peprp(x_2,b_2)
 +(x_3-x_2)\peprt(x_2,b_2) \right ] \pep(x_3,b_2)
\right.\right.\non
 & & \left.\left.
  +\re \rep \left [ (x_3-x_2)\peprp(x_3,b_2) +(x_2+x_3-2)\peprt(x_2,b_2)\right
  ]\pet(x_3,b_2)
\right\} \right. \non && \left. \quad \cdot
\alpha_s(t_f^3)h_f^3(x_1,x_2,x_3,b_1,b_2)\exp[-S_{ef}(t_f^3)]
\right.  \non && \left.
 -\left \{ x_2\pe(x_3,b_2) \pepr(x_2,b_2)
\right.\right. \non
 && \left. \left.
 +\re \rep
 \left [ (x_2+x_3)\peprp(x_2,b_2)+(x_2-x_3)\peprt(x_2,b_2) \right ]\pep(x_3,b_2)
\right. \right.\non && \left.\left.
 +\re \rep
 \left [(x_2-x_3)\peprp(x_2,b_2)+(x_2+x_3)\peprt(x_2,b_2) \right ]\pet(x_3,b_2)
 \right \}
\right. \non && \left.
 \quad \cdot  \alpha_s(t_f^4)h_f^4(x_1,x_2,x_3,b_1,b_2)\exp[-S_{ef}(t_f^4) ]
 \right \}\; . \label{eq:mapip2}
\eeq

In the above equations, we have assumed that $x_1 <<x_2,x_3$. Since
the light quark momentum fraction $x_1$ in $B$ meson is peaked at
the small region, while quark momentum fraction $x_3$ of $\eta$ is
peaked around $0.5$, this is not a bad approximation. The numerical
results also show that this approximation makes very little
difference in the final result. After using this approximation, all
the diagrams are functions of $k_1^+= x_1 m_B/\sqrt{2}$ of B meson
only, independent of the variable of $k_1^-$. Therefore the
integration of eq.(\ref{int}) is performed safely.

For the $B \to \eta\eta^{\prime}$ decay, besides the Feynman diagrams as shown in Fig.~1 where
the upper emitted meson is the $\etapr$, the Feynman diagrams obtained by exchanging
the position of $\eta$ and $\etapr$ also contribute to this decay mode.
The corresponding expressions of amplitudes for new diagrams will be similar with those as given
in Eqs.(\ref{eq:ab}-\ref{eq:mapip2}), since the $\eta$ and $\etapr$ are
all light pseudoscalar mesons and have the similar wave functions. The
expressions of amplitudes for new diagrams can be obtained by the
replacements
\beq
\pe \longleftrightarrow  \phi_{\etapr},   \quad
\pep\longleftrightarrow  \phi^P_{\etapr}, \quad
\pet \longleftrightarrow \phi^t_{\etapr}, \quad
\re  \longleftrightarrow  r_{\etapr}.
\eeq

 For example, we find that:
 \beq
  F_{e\etapr}&=&F_{e\eta},\quad F_{a\etapr}=-F_{a\eta},
  \quad F_{a\etapr}^{P1}=-F_{a\eta}^{P1},\quad
 F_{a\etapr}^{P2}=F_{a\eta}^{P2}.
 \eeq

%%%%%%%%%%%%%%%%%%%%%%%%%%%%%%%%%%%%%%%%%%%%%%%%%%%%%%%%%%%%%%%%%

\subsection{Mixing of $\eta$ and $\eta^\prime$ meson}

Before we write down the complete decay amplitude for the studied
decay modes, we firstly give a brief discussion about the
$\eta-\eta^\prime$ mixing and the gluonic component of the $\eta^\prime$ meson.
There exist two popular mixing basis for $\eta-\eta^\prime$ system, the octet-singlet and the
quark flavor basis, in literature.
Here we  use the $SU(3)_F$ octet-singlet basis with the two mixing angle scheme \cite{lk98,fk98}
instead of the simple one mixing angle
scheme to describe the mixing of $\eta$ and $\eta^\prime$ mesons, since the former scheme
can archive a better agreement with the relevant data, such as the decay width of
$\etap\to \gamma\gamma$, the $\etap\gamma$ transition form factors,
the radiative $J/\Psi$ decays\cite{fk98,0501072}.
In the two mixing angle scheme, the meson $\eta$, $\eta^\prime$ and the decay constants
can be defined as
\beq
\left( \begin{array}{c}
\eta\\ \eta^\prime \\ \end{array} \right )
&=& \left( \begin{array}{cc}
 \cos{\theta_8} & -\sin{\theta_1} \\
 \sin{\theta_8} & \cos\theta_1 \end{array} \right )
 \left( \begin{array}{c}  \eta_8\\ \eta_1 \\ \end{array} \right ), \label{eq:e-ep}
\eeq
with  the flavor $SU(3)$-octet and -singlet components
\beq
\eta_1 = ( u\bar{u}+d\bar{d}+s\bar{s})/\sqrt{3}, \quad
\eta_8 = \left ( u\bar{u}+d\bar{d}-2s\bar{s}\right )/\sqrt{6},
\label{eq:eta18}
\eeq
in the quark model, and the two mixing angles $\theta_1$ and $\theta_8$, in principle, can be
determined by various related experiments \cite{fk98,0501072}.
In the numerical calculations, we will use the following $\eta-\eta^\prime$ mixing parameters
\cite{fk98}
\beq
\theta_8 &=&-21.2^\circ, \quad \theta_1=-2.4^\circ, \non
f_1&=& 151 {\rm MeV}, \quad f_8 = 169 {\rm MeV},   \label{eq:t1-t8}
\eeq
obtained by setting $f_q = f_\pi=130$ MeV and $f_s = \sqrt{2f_K^2 -f_\pi^2} = 1.41 f_\pi$ \cite{fk98}.
The second set of the $\eta-\eta^\prime$ mixing parameters obtained phenomenological and used frequently
in literature reads \cite{fk98}
\beq
\theta_8 &=&-21.2^\circ, \quad \theta_1=-9.2^\circ, \non
f_1&=& 1.17 f_\pi = 152 {\rm MeV}, \quad f_8 = 1.26 f_\pi = 164 {\rm MeV},   \label{eq:mp2}
\eeq
for $f_\pi=130 MeV$. We usually use the first set of mixing parameters in numerical calculations,
unless otherwise stated.

As shown in Eq.~(\ref{eq:eta18}), $\eta$ and $\eta^\prime$ are generally considered as a linear
combination of light quark pairs $u\bar{u}, d\bar{d}$ and $s\bar{s}$.
But it should be noted that the $\eta^\prime$ meson may have a gluonic
component in order to interpret the anomalously large branching ratios of $B\to K \eta^\prime$
\cite{ekou1,ekou2,bn03,li0609}.
Although some progress have been achieved in recent years about this problem,
we currently still do not know how to calculate reliably the gluonic contributions to the
B meson two body decay modes involving $\eta^\prime$ meson.
For the studied decay modes in this paper, on the other hand, currently available measurements
for the branching ratios still have big uncertainties. It is therefore reasonable at present to consider
only the dominant contributions from the quark contents of $\eta^\prime$ meson, while take the
subdominant contribution from the possible gluonic content of $\eta^\prime$ meson as a source of
theoretical uncertainties. Following Ref.~\cite{li0609}, on the other hand,
 we also estimated the possible
gluonic contributions to $B \to \etap \etap$ decays induced by the gluonic corrections to the $B \to \etap$
transition form factors \cite{li0609} and found that these corrections to both the branching ratios
and CP violating asymmetries are indeed very small.

%%%%%%%%%%%%%%%%%%%%%%%%%%%%%%%%%%%%%%%%%%%%%%%%%%%%%%%%%%%%%%%

\subsection{Complete decay amplitudes}

For $B^0 \to \eta \eta$ decay, by combining the contributions from different diagrams,
the total decay amplitude can be written as
\beq
{\cal M}(\eta\eta) &=&
\sqrt{2}\left \{F_{e\eta} F_1(\theta_1,\theta_8) \left \{\left[ \xi_u \left(
C_1 + \frac{1}{3}C_2\right) \right.\right.\right.
 \non
& &\left.\left.\left.-\xi_t
\left(\frac{7}{3}C_3+\frac{5}{3}C_4-2C_{5}-\frac{2}{3}C_{6}
-\frac{1}{2}C_7-\frac{1}{6}C_8+\frac{1}{3}C_9
 -\frac{1}{3} C_{10}\right)\right ] f_\eta^d \right.\right.\non
& &\left.\left.-
\xi_t\left(C_{3}+\frac{1}{3}C_{4}-C_{5}-\frac{1}{3}C_{6}+
\frac{1}{2}C_7+\frac{1}{6}C_8-\frac{1}{2}C_9-\frac{1}{6}C_{10}\right)
f_\eta^s\right \} \right.\non && \left.-
F_{e\eta}^{P_2} F_1(\theta_1,\theta_8) \xi_t \left (\frac{1}{3}C_5+C_6
-\frac{1}{6}C_7-\frac{1}{2}C_{8}\right) f_{\eta}^d \right.\non
&&\left. + M_{e\eta} F_1(\theta_1,\theta_8) \left \{ \left [ \xi_uC_2-\xi_t
\left(C_3+2C_4-\frac{1}{2}C_9 +\frac{1}{2}C_{10}\right)\right ] F_1(\theta_1,\theta_8)
 \right.\right. \non && \left.\left. -\xi_t \left ( C_4
-\frac{1}{2}C_{10}\right ) F_2(\theta_1,\theta_8) \right \} \right.\non &&
\left.- M_{e\eta}^{P_2} F_1(\theta_1,\theta_8) \, \xi_t\,
\left[\left(2C_6+\frac{1}{2}C_8\right) F_1(\theta_1,\theta_8)
+\left(C_6-\frac{1}{2}C_8\right)F_2(\theta_1,\theta_8) \right]\right. \non
&&\left. + M_{a\eta}\left \{\left[ \xi_uC_{2}-\xi_t
\left(C_3+2C_4-\frac{1}{2}C_9 +\frac{1}{2}C_{10}\right)\right
]F_1(\theta_1,\theta_8)^{2}\right. \right.\non && \left. \left.-\xi_t \left
( C_4 -\frac{1}{2}C_{10}\right ) F_2(\theta_1,\theta_8)^{2}\right
\}-M_{a\eta}^{P_1}\,
\xi_t\,\left(C_{5}-\frac{1}{2}C_{7}\right) F_1(\theta_1,\theta_8)^{2}
 \right.\non && \left.-
M_{a\eta}^{P_2}\, \xi_t\,
\left[\left(2C_6+\frac{1}{2}C_8\right) F_1(\theta_1,\theta_8)^{2}
+\left(C_6-\frac{1}{2}C_8\right) F_2(\theta_1,\theta_8)^{2}\right]
\right.\non &&\left.-F_{a\eta}^{P2}\,\xi_t \left (\frac{1}{3}C_5+C_6
-\frac{1}{6}C_7-\frac{1}{2}C_{8}\right) F_1(\theta_1,\theta_8)^{2}\cdot
f_{B} \right\} , \label{eq:m1}
\eeq
where $\xi_u = V_{ub}^*V_{ud}$, $\xi_t = V_{tb}^*V_{td}$, while the mixing parameters and the relevant
decay constants are
\beq
F_1(\theta_1,\theta_8)&=& \sqrt{\frac{1}{6}}\cos\theta_8 - \sqrt{\frac{1}{3}} \sin\theta_1, \quad
F_2(\theta_1,\theta_8)=-\sqrt{\frac{2}{3}}\sin\theta_8 + \sqrt{\frac{1}{3}} \cos\theta_1,\label{eq:f1f2} \\
f_{\eta}^d &=&  \frac{f_8}{\sqrt{6}} \cos\theta_8 - \frac{f_1}{\sqrt{3}}\sin\theta_1,
\quad
f_{\eta}^s = -\frac{2 f_8}{\sqrt{3}} \cos\theta_8  + \frac{f_1}{\sqrt{3}} \sin\theta_1.
\eeq
It should be mentioned that the Wilson coefficients $C_i=C_i(t)$
in Eq.~(\ref{eq:m1}) should be calculated at the appropriate scale $t$
using equations as given in the Appendices of Ref.~\cite{luy01}. Here the scale $t$ in the Wilson
coefficients should be taken as the same scale appeared in the expressions of decay amplitudes
from Eqs.~(\ref{eq:ab}) to (\ref{eq:mapip2}). This is the way in pQCD approach to eliminate
the scale dependence.

Similarly, the decay amplitude for $B^0 \to \eta \etapr$ can be written as
\beq
{\cal M}(\eta\etapr) &=& \left
 (F_{e\eta}F_1(\theta_1,\theta_8)f_{\eta^\prime}^d +F_{e\etapr}F'_1(\theta_1,\theta_8)f_\eta^d\right )
 \cdot\left[ \xi_u \left( C_1 +
\frac{1}{3}C_2\right)\right.
 \non
& &\left.-\xi_t
\left(\frac{7}{3}C_3+\frac{5}{3}C_4-2C_{5}-\frac{2}{3}C_{6}
-\frac{1}{2}C_7-\frac{1}{6}C_8+\frac{1}{3}C_9
 -\frac{1}{3} C_{10}\right)\right ]
 \non
& &- \left( F_{e\eta}F_1(\theta_1,\theta_8)f_{\eta^\prime}^s
+F_{e\etapr}F'_1(\theta_1,\theta_8)f_\eta^s\right )\non & &
\cdot\,\xi_t\left(C_{3}+\frac{1}{3}C_{4}-C_{5}-\frac{1}{3}C_{6}+
\frac{1}{2}C_7+\frac{1}{6}C_8-\frac{1}{2}C_9-\frac{1}{6}C_{10}\right)
\non && -
 \left(F_{e\eta}^{P_2}F_1(\theta_1,\theta_8)f_{\eta^{\prime}}^d
+F_{e\etapr}^{P_2}F'_1(\theta_1,\theta_8)f_{\eta}^d\right)\xi_t \left
(\frac{1}{3}C_5+C_6 -\frac{1}{6}C_7-\frac{1}{2}C_{8}\right) \non &&
+\left( M_{e\eta}+
M_{e\etapr}\right)F_1(\theta_1,\theta_8)F'_1(\theta_1,\theta_8)\left [ \xi_uC_2-\xi_t
\left(C_3+2C_4-\frac{1}{2}C_9 +\frac{1}{2}C_{10}\right)\right ]
  \non &&  -\left[ M_{e\eta}F_1(\theta_1,\theta_8)F_2'(\theta_1,\theta_8)+
M_{e\etapr}F'_1(\theta_1,\theta_8)F_2(\theta_1,\theta_8)\right] \xi_t \left ( C_4
-\frac{1}{2}C_{10}\right )
  \non && -\left( M_{e\eta}^{P_2}+
  M_{e\etapr}^{P_2}\right)\,F_1(\theta_1,\theta_8)F'_1(\theta_1,\theta_8)
\xi_t\,\left(2C_6+\frac{1}{2}C_8\right)\non && -
\left(M_{e\eta}^{P_2}F_1(\theta_1,\theta_8)F_2'(\theta_1,\theta_8)+
M_{e\etapr}^{P_2}F'_1(\theta_1,\theta_8)F_2(\theta_1,\theta_8)\right)\xi_t\,
\left(C_6-\frac{1}{2}C_8\right)\non && +
\left(M_{a\eta}+M_{a\etapr}\right) F_1(\theta_1,\theta_8)F'_1(\theta_1,\theta_8)\left[
\xi_uC_{2}-\xi_t \left(C_3+2C_4-\frac{1}{2}C_9
+\frac{1}{2}C_{10}\right)\right ] \non &&
 -\left(M_{a\eta}+M_{a\etapr}\right)F_2(\theta_1,\theta_8)F'_2(\theta_1,\theta_8)\xi_t
\left ( C_4 -\frac{1}{2}C_{10}\right ) \non &&
-\left(M_{a\eta}^{P_1}\,+M_{a\etapr}^{P_1}\,\right)F_1(\theta_1,\theta_8)F'_1(\theta_1,\theta_8)
\xi_t\,\left(C_{5}-\frac{1}{2}C_{7}\right)
 \non && -
 \left(M_{a\eta}^{P_2}\,+
 M_{a\etapr}^{P_2}\,\right) F_1(\theta_1,\theta_8)F'_1(\theta_1,\theta_8)\xi_t\,
\left(2C_6+\frac{1}{2}C_8\right) \non && -
 \left(M_{a\eta}^{P_2}\,+
 M_{a\etapr}^{P_2}\,\right)F_2(\theta_1,\theta_8)F'_2(\theta_1,\theta_8)\xi_t\,
 \left(C_6-\frac{1}{2}C_8\right)
\non &&-f_{B}\cdot\left(F_{a\eta}^{P2}+
F_{a\etapr}^{P2}\right)F_1(\theta_1,\theta_8)F'_1(\theta_1,\theta_8)\xi_t \left
(\frac{1}{3}C_5+C_6 -\frac{1}{6}C_7-\frac{1}{2}C_{8}\right)
, \label{eq:m2}
\eeq
where the relevant mixing parameters and decay constants are
\beq
F'_1(\theta_1,\theta_8) &=& \sqrt{\frac{1}{6}} \sin{\theta_8} + \sqrt{\frac{1}{3}} \cos{\theta_1}, \quad
F'_2(\theta_1,\theta_8) = -\sqrt{\frac{2}{3}} \sin{\theta_8}  + \sqrt{\frac{1}{3}} \cos{\theta_1}, \\
f_{\eta^\prime}^d &=& \frac{f_8}{\sqrt{6}} \sin\theta_8 + \frac{f_1}{\sqrt{3}}\cos\theta_1,
\quad
f_{\eta^\prime}^s = -\frac{2 f_8}{\sqrt{3}} \sin\theta_8 + \frac{f_1}{\sqrt{3}} \cos\theta_1 .
\eeq

The complete decay amplitude  ${\cal M}(\etapr \etapr)$ for $B \to \eta^{\prime} \eta^{\prime}$
decay  can be obtained easily from Eq.(\ref{eq:m1})  by the following
replacements
\beq
f_\eta^{d},\; f_\eta^s &\longrightarrow & f_{\eta^\prime}^d, \; f_{\eta^\prime}^s, \non
F_1(\theta_1,\theta_8) &\longrightarrow & F'_1(\theta_1,\theta_8), \quad
F_2(\theta_1,\theta_8) \longrightarrow  F'_2(\theta_1,\theta_8).
\eeq
Note that the contributions from the possible gluonic component of $\eta'$ meson have not
been included here.

\section{Numerical results and Discussions}\label{sec:n-d}

In this section, we will calculate the branching ratios and CP violating asymmetries
for those considered decay modes. The input parameters and the wave functions to be used are given
in Appendix \ref{sec:app2}. In numerical calculations, central values of input parameters will be used
implicitly unless otherwise stated.

Based on the definition of the form factor $F_0^{B \to \eta}$ as given in Eq.~(\ref{eq:f01}) and consider
the relation of $F_{e\eta} = F_{e\etapr}$ at the leading order of pQCD approach,
we find the numerical values of the
corresponding form factors at zero momentum transfer:
\beq
F_{0,1}^{B \to \eta}(q^2=0)= F_{0,1}^{B \to \eta^{\prime}}(q^2=0)= \frac{F_{e\eta}}{m_B^2}
=0.30^{+0.05}_{-0.04}(\omega_b),
\label{eq:aff0}
\eeq
for $\omega_b=0.40 \pm 0.04$ GeV, which agrees well with those as given in Refs.~\cite{ball,pball}.

\subsection{Branching ratios}

Using the decay amplitudes obtained in last section, it is straightforward to calculate the branching ratios.
For $B^0 \to \eta \eta, \eta \etapr$ and $\etapr \etapr $ decays, the decay amplitudes as given  in
Eqs.~(\ref{eq:m1}) and (\ref{eq:m2}) can be rewritten as
 \beq
{\cal M} &=& V_{ub}^*V_{ud} T -V_{tb}^* V_{td} P= V_{ub}^*V_{ud} T
\left [ 1 + z e^{ i ( \alpha + \delta ) } \right], \label{eq:ma}
\eeq
where
\beq
z=\left|\frac{V_{tb}^* V_{td}}{ V_{ub}^*V_{ud} }\right| \left|\frac{P}{T}\right| \label{eq:zz}
\eeq
is the ratio of penguin to tree contributions,
$\alpha = \arg \left[-\frac{V_{td}V_{tb}^*}{V_{ud}V_{ub}^*}\right]$ is the weak
phase (one of the three CKM angles), and $\delta$ is the relative
strong phase between tree (T) and penguin (P) diagrams \footnote{The ``T" and ``P" term refers to the part of
the decay amplitude ${\cal M}$ in Eq.~(\ref{eq:ma}), which is
proportional to $V_{ub}^*V_{ud} $ and $V_{tb}^*V_{td}$ respectively.}.
In the pQCD approach, the ratio $z$ and the
strong phase $\delta$ can be calculated perturbatively.

From Eq.~(\ref{eq:ma}), it is easy to write the decay amplitude for the
corresponding charge conjugated decay mode
\beq
\overline{\cal M} &=& V_{ub}V_{ud}^* T -V_{tb} V_{td}^* P = V_{ub}V_{ud}^* T \left[1
+z e^{i(-\alpha + \delta)} \right]. \label{eq:mb}
 \eeq
Therefore the CP-averaged branching ratio for $B^0 \to \etap \etap$ decays is
\beq
Br = (|{\cal M}|^2 +|\overline{\cal M}|^2)/2 =  \left|
V_{ub}V_{ud}^* T \right| ^2 \left[1 +2 z\cos \alpha \cos \delta +z^2
\right], \label{br}
\eeq
where the ratio $z$ and the strong phase $\delta$ have been defined in Eqs.(\ref{eq:ma}) and (\ref{eq:zz}).

By employing the two mixing angle scheme of $\eta-\eta^\prime$ system
and using the mixing parameters as given in Eq.~(\ref{eq:t1-t8}), one finds the CP averaged
branching ratios for the considered three decays as follows
\beq
Br(\ B^0 \to\eta \eta) &=& \left [1.5\pm 0.3(\omega_b)
 ^{+0.1}_{-0.2}(m_0^\pi)\pm0.4(\alpha )\right ] \times 10^{-7}, \label{eq:bree1}\\
Br(\ B^0 \to \eta\eta^{\prime}) &=& \left [0.60^{+0.19} _{-0.14}
( \omega_b) \pm0.05 (m_0^\pi)\pm0.04 ( \alpha ) \right ] \times 10^{-7}, \label{eq:brep1} \\
Br(\ B^0 \to \eta^{\prime}\eta^{\prime}) &=& \left [0.68 ^{+0.08}
_{-0.09} (\omega_b) \pm 0.05 (m_0^\pi)^{+0.22} _{-0.19} (\alpha ) \right ]\times 10^{-7},
 \label{eq:brpp1}
\eeq
where the main errors are induced by the uncertainties of $\omega_b=0.4 \pm 0.04$ GeV,
 $m_0^\pi = 1.4 \pm 0.1$ GeV and $\alpha =100^\circ \pm 20^\circ$, respectively.

Besides the theoretical uncertainties as shown in Eqs.~(\ref{eq:bree1}-\ref{eq:brpp1}), the change of
the mixing scheme of $\eta-\eta^\prime$ system or changes of mixing parameters within a
given scheme may also induce  moderate or even significant changes to the theoretical predictions.
The central values of the branching ratios for the considered three decays, for example, will be
\beq
Br(\ B^0 \to\eta \eta)\left |_{CV} \right.&=& 1.9\times 10^{-7}, \label{eq:br21}\\
Br(\ B^0 \to \eta\eta^{\prime})\left |_{CV} \right. &=& 0.82  \times 10^{-7}, \label{eq:br22} \\
Br(\ B^0 \to \eta^{\prime}\eta^{\prime})\left |_{CV} \right. &=& 0.66 \times 10^{-7},
\label{eq:br23}
\eeq
if the second set of mixing parameters in the two mixing angle scheme, as given in Eq.~(\ref{eq:mp2}),
are used, and
\beq
Br(\ B^0 \to\eta \eta)\left |_{CV} \right. &=& 1.9\times 10^{-7}, \label{eq:br31}\\
Br(\ B^0 \to \eta\eta^{\prime})\left |_{CV} \right. &=& 1.2  \times 10^{-7}, \label{eq:br32} \\
Br(\ B^0 \to \eta^{\prime}\eta^{\prime})\left |_{CV} \right. &=& 0.69 \times 10^{-7},
\label{eq:br33}
\eeq
if the one mixing angle scheme with $\theta_p=-12.3^\circ$ are employed. From the above
numerical results one can see that (a) the induced variation is  about $2\%$ or $20\%$ for
$B \to \eta^\prime \eta^\prime$ and $B \to \eta \eta$ decay respectively, which is consistent
with the general expectation; (b) for $B \to \eta \eta^\prime$ decay, however,
the resultant change of the branching ratio is roughly a factor of two.
Such variation is channel-dependent and should be considered as one kind of theoretical uncertainties.
The large change in $Br(B \to \eta \eta^\prime)$ is induced by destructive
interference between the individual decay amplitudes
from different Feynman diagrams when two mixing angle scheme are employed.

As a comparison, furthermore, we also list here the theoretical predictions for the
branching ratios of the three
decays in the QCDF approach\cite{bn03b} and the SCET approach
\footnote{The theoretical predictions for the branching ratio and the direct CP violating asymmetries
in the SCET approach, as shown in Eqs.~(\ref{eq:br11}-\ref{eq:br24}) and Eqs.(\ref{eq:acp11}-\ref{eq:acp24}),
are directly quoted from Table VII of Ref.~\cite{wz0601}, for the case of Theory II. } \cite{wz0601}:
\beq
 Br(\ B^0 \to\eta \eta) &=&  \left\{ \begin{array}{ll}
\left ( 0.16 ^{+0.45}_{-0.19}\right ) \times 10^{-6}, &\ \ \ { \rm QCDF}, \\
\left ( 1.0 \pm 1.5 \right ) \times 10^{-6}, & \ \ \ {\rm SCET}, \\
\end{array} \right. \label{eq:br11}\\
 Br(\ B^0 \to \eta\eta^{\prime}) &=& \left\{ \begin{array}{ll}
\left ( 0.16 ^{+0.61}_{-0.18}\right ) \times 10^{-6}, & \ \ \ {\rm QCDF}, \\
\left ( 2.2 \pm 5.5 \right ) \times 10^{-6}, & \ \ \ {\rm SCET}, \\
\end{array} \right. \\
 Br(\ B^0 \to\eta^{\prime} \eta^{\prime}) &=& \left\{ \begin{array}{ll}
\left ( 0.06 ^{+0.25}_{-0.07}\right ) \times 10^{-6}, &\ \ \  {\rm QCDF}, \\
\left ( 1.2 \pm 3.8 \right ) \times 10^{-6}, &\ \ \  {\rm SCET}, \\
\end{array} \right.
 \label{eq:br24}
\eeq
where the individual errors as given in Refs.~\cite{bn03b} and \cite{wz0601} have been added in
quadrature. It is easy to see that (a) the theoretical predictions for
$Br(B \to \eta \eta)$ and $Br(B \to \eta^\prime \eta^\prime)$ in both the QCDF and the pQCD approaches
agree very well; (b) for $Br(B \to \eta \eta^\prime)$, the central value of the pQCD prediction is
about half of the QCDF prediction, but still agree within one standard deviation; and  (c) the
central values of the theoretical predictions in the SCET approach are much larger than those in QCDF and
pQCD approaches, but still consistent with them if one takes the very large theoretical
uncertainties in SCET approach into account.

It is worth of mentioning that the FSI effects are not considered here.
The smallness of FSI effects for B meson decays into two  light mesons
has been put forward by Bjorken \cite{b89}
based on the color transparency argument \cite{lb80}, and also supported by further
renormalization group analysis of soft gluon exchanges among initial and final state mesons \cite{soft}.

\begin{figure}[tb]
\centerline{\mbox{\epsfxsize=9cm\epsffile{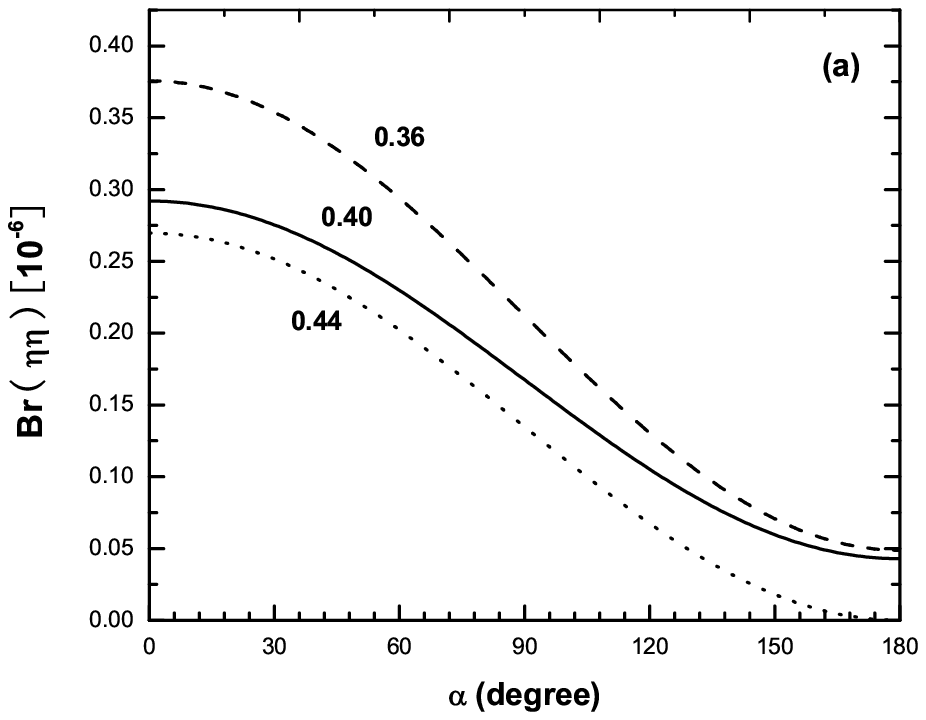}\epsfxsize=9cm\epsffile{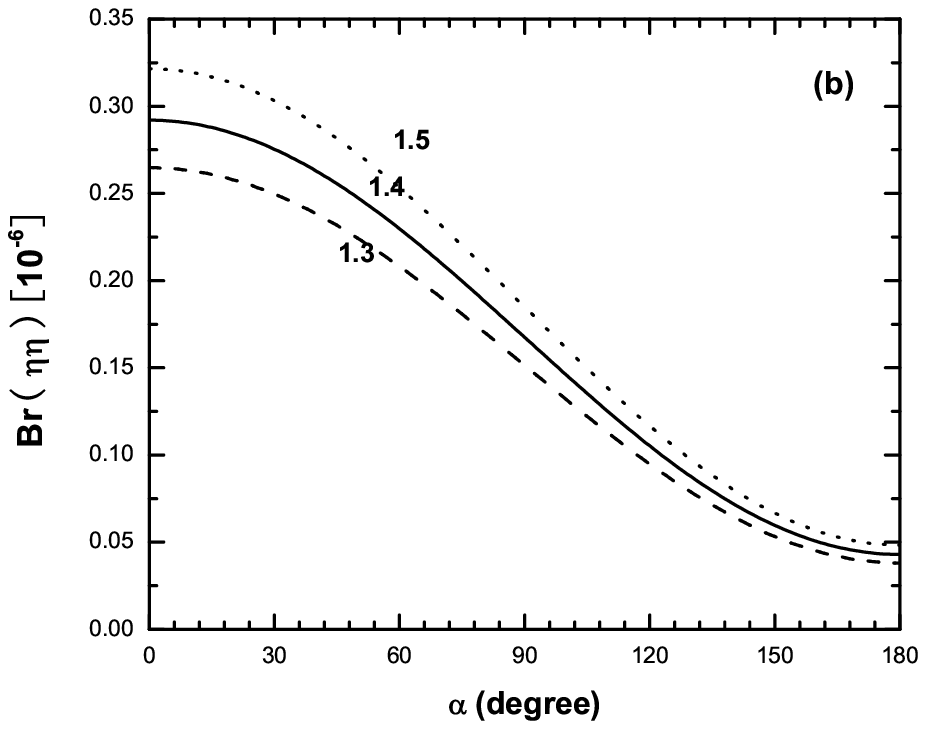}}}
\vspace{0.3cm}
\caption{The $\alpha$ dependence of the branching ratios (in units of
 $10^{-6}$)  of $B^0\to \eta \eta $ decays. Here (a) is for $m_0^\pi=1.4$ GeV,
 $\omega_b= 0.40\pm 0.04 $ GeV; and (b) is for $\omega_b=0.4$ GeV,   $m_0^\pi=1.4\pm 0.1$ GeV. }
\label{fig:fig2}
\end{figure}

\begin{figure}[tb]
\centerline{\mbox{\epsfxsize=9cm\epsffile{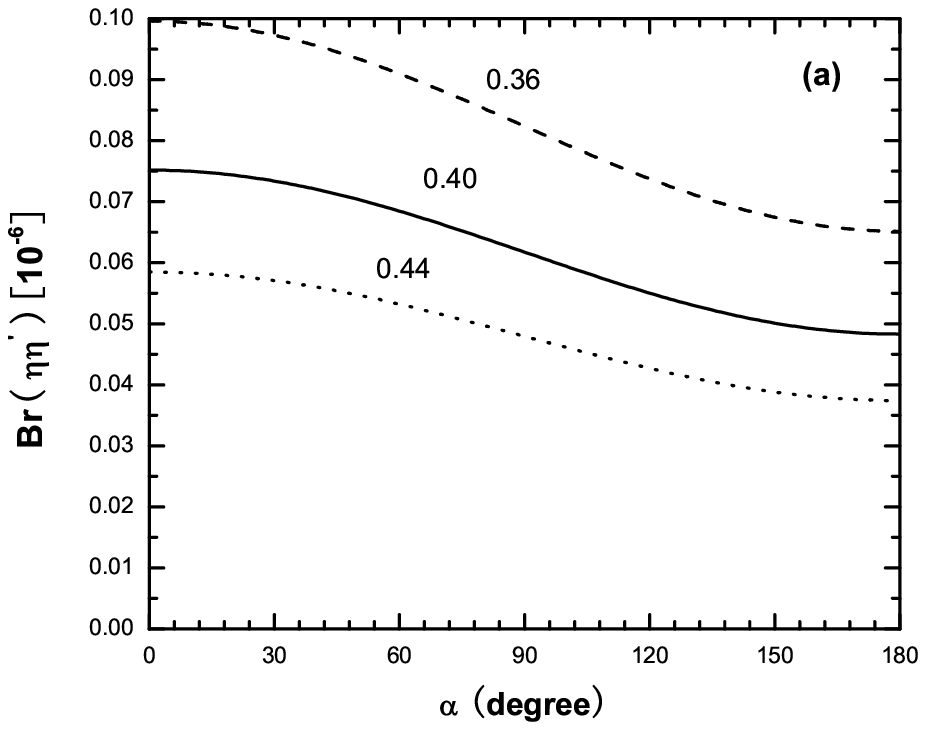}\epsfxsize=9cm\epsffile{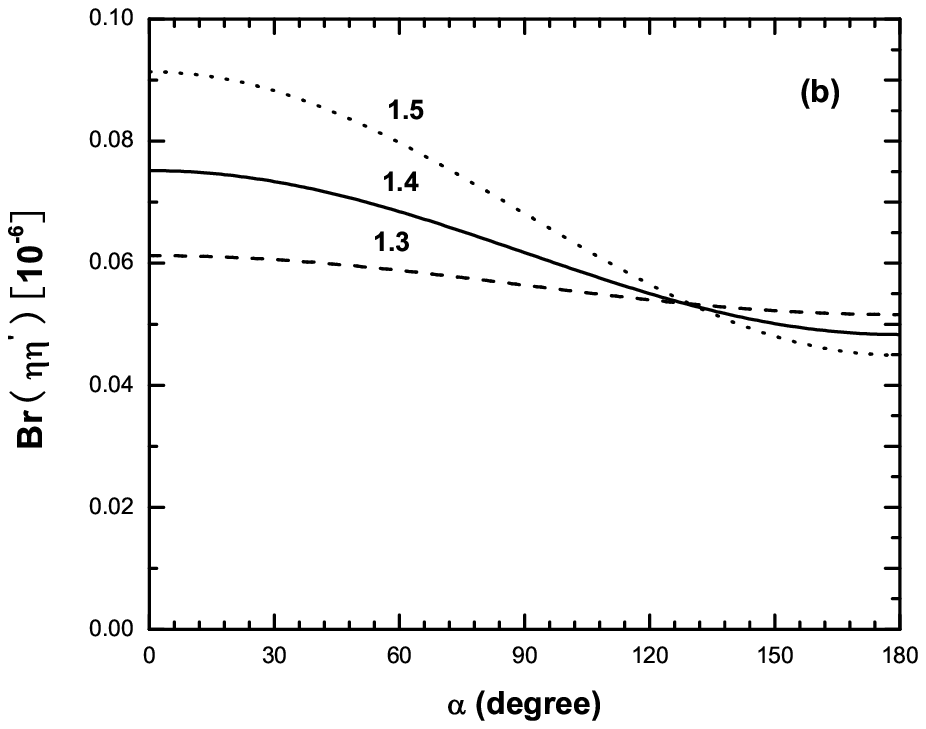}}}
\vspace{0.3cm}
 \caption{The same as Fig.~\ref{fig:fig2} but for  $B^0\to \eta \eta^{\prime} $ decay.}
 \label{fig:fig3}
\end{figure}

\begin{figure}[tb]
\centerline{\mbox{\epsfxsize=9cm\epsffile{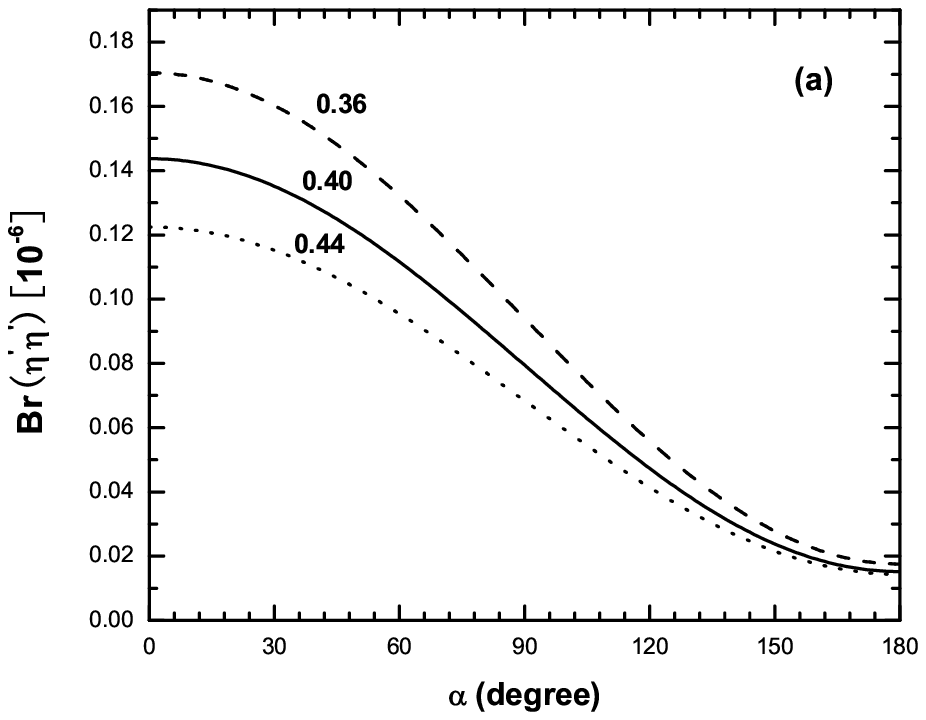}\epsfxsize=9cm\epsffile{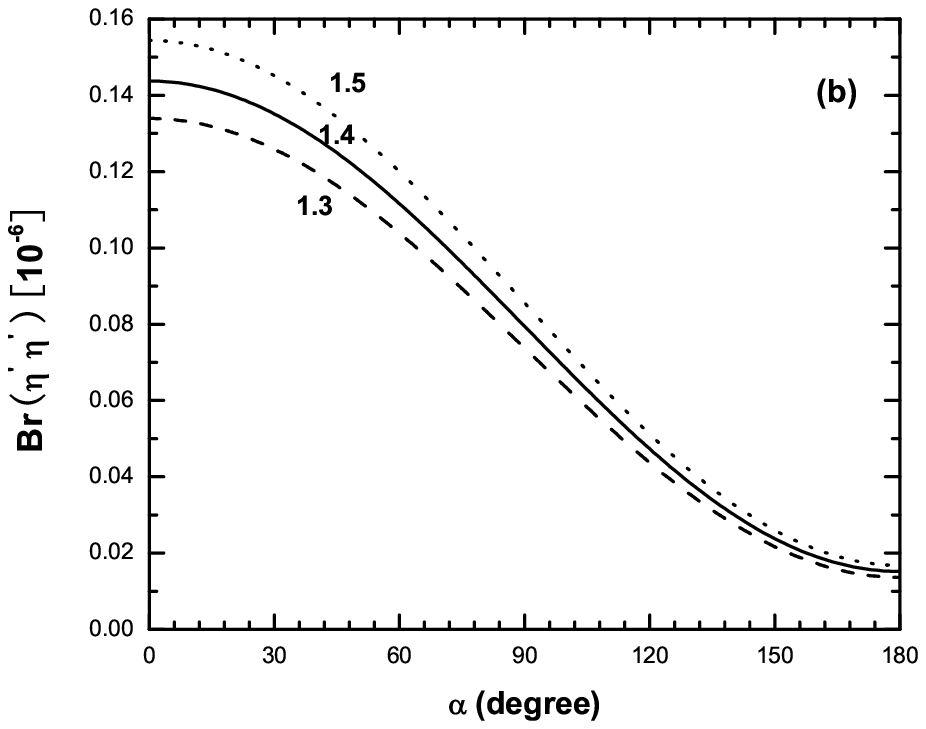}}}
\vspace{0.3cm}
 \caption{The same as Fig.~\ref{fig:fig2} but for $B^0\to \eta^{\prime}\eta^{\prime} $ decay.}
 \label{fig:fig4}
\end{figure}

At present, the pQCD predictions still have large theoretical errors
( say $\sim 50\%$ ) induced by the large uncertainties of many input parameters.
In our analysis, we considered the constraints on these parameters from analysis of other well
measured decay channels. For example, the constraint $ 1.1
\mbox{GeV} \leq m_0^\pi \leq 1.9 \mbox{GeV}$ was obtained from the
phenomenological studies for $B \to \pi \pi$ decays \cite{luy01},
while the constraint  of $\alpha = \left ( 93^{+ 11}_{-9} \right )^\circ$ are obtained
from the direct measurements \cite{hazumi06}.
In estimating the uncertainties, we still conservatively consider the range of
$\alpha = 100^\circ \pm 20^\circ$.

In Figs.~\ref{fig:fig2}, ~\ref{fig:fig3} and \ref{fig:fig4}, we
show the parameter dependence of the pQCD predictions for the branching ratios
of $B \to \eta\eta$ , $ \eta\etapr$ and $\etapr \etapr$ decays for
$\omega_b=0.4\pm 0.04$ GeV, $m_0^\pi=1.4\pm 0.1$ GeV and $\alpha=[0^\circ,180^\circ]$.
From the numerical results, we observe that the pQCD predictions are sensitive to the
variations of $\omega_b$, $m_0^\pi$ and $\alpha$.

\subsection{Branching ratios with updated wave functions}

In pQCD  approach, the only input matters are wave functions of the involved particles,
which stand for nonperturbative contributions. Since currently available wave functions are not
exactly determined, large error of the theoretical predictions will be produced from
the uncertainty of the relevant $B$, $\eta$ and $\eta^\prime$ wave functions
\footnote{This constitutes a large theoretical uncertainty common to the pQCD, QCDF and other
factorization approaches where the meson wave functions are used as input. }.

In last subsection, we choose the pion DAs derived from QCD sum rule
as given in Refs.~\cite{ball,kf} with a fixed decay constant
$f_{B}=190$ MeV(see Appendix \ref{sec:app2} for more details). The
resultant pQCD predictions for CP averaged branching ratios are
consistent with the measured values, which may be regarded as an
indication that above inputs are reasonable.

In this subsection, in order to check the theoretical uncertainty induced by the variation of wave functions,
we  recalculate the CP averaged branching ratios by employing the updated models
of the pion DAs as given in Ref.~\cite{pball}. Although the structure of
$\phi_\pi^{A,P,T}$ as given in Appendix \ref{sec:app2}
remain unchanged, but the Gegenbauer moments $a_2^\pi$ and $a_4^\pi$ in the updated pion DAs
are changed significantly, they are now
\beq
\left (  a_{2}^{\pi},  a_{4}^{\pi} \right ) &=& \left ( 0.115, -0.015 \right ),
\label{eq:a2pin}
 \eeq
instead of the old $(0.44,0.25)$ as shown in Eq.~(\ref{eq:a2pi}).

Using the updated pion DAs with  $( a_{2}^{\pi},  a_{4}^{\pi} ) = ( 0.115, -0.015)$ and
$f_B=210$ MeV \cite{li05b}, we find numerically that

\begin{enumerate}
\item[]{(i)}
The values of the form factors as given in Eq.~(\ref{eq:aff0}) will be decreased by about $10\%$, and
we now have
\beq
F_{0,1}^{B \to \eta}(q^2=0)= F_{0,1}^{B \to \eta^{\prime}}(q^2=0)= 0.27^{+0.05}_{-0.03}(\omega_b),
\label{eq:aff1}
\eeq
for $\omega_b=0.40\pm 0.04$ GeV.

\item[]{(ii)}
The CP averaged branching ratios of the considered decays are
\beq
 Br(\ B^0 \to\eta \eta) &=& \left [0.67^{+0.22}_{-0.15}(\omega_b)
 ^{+0.09}_{-0.08} (m_0^\pi)^{+0.14}_{-0.13}(\alpha )
 ^{+0.15}_{-0.12}(a_2^\pi)\right ] \times 10^{-7}, \label{eq:bree3}\\
Br(\ B^0 \to \eta\eta^{\prime}) &=& \left [0.18^{+0.05} _{-0.03}
( \omega_b) \pm0.01 (m_0^\pi)\pm 0.01(\alpha )^{+0.10}_{-0.08}(a_{2}^\pi)
\right ] \times 10^{-7}, \label{eq:brep3} \\
Br(\ B^0 \to \eta^{\prime}\eta^{\prime}) &=& \left [0.11 ^{+0.04}_{-0.01} (\omega_b)
\pm 0.01 (m_0^\pi)\pm0.04 (\alpha) ^{+0.10}_{-0.07} (a_{2}^\pi) \right ]\times 10^{-7},  \label{eq:brpp3}
\eeq
where the main errors are induced by the uncertainties of $\omega_b=0.4 \pm 0.04$ GeV,
$m_0^\pi=1.4 \pm 0.1$ GeV, $\alpha =100^\circ \pm 20^\circ$, $a_2^\pi=0.115 \pm 0.115$, respectively.
The errors induced by varying $a_4^\pi$ in the range of $a_4^\pi=-0.015 \pm 0.015$ are very small:
only about $\pm 0.03, \pm 0.004$ and $\pm 0.01$ (in unit of $10^{-7}$) for $B \to \eta\eta, \eta \eta^\prime$
and $\eta^\prime \eta^\prime$ decay, respectively.
We here have assumed that the updated Gegenbauer moments $a_2^\pi$ and $a_4^\pi$ can vary by $100\%$.

\end{enumerate}

It is easy to see that the pQCD predictions in Eqs.(\ref{eq:bree3}-\ref{eq:brpp3})
is smaller than those in Eqs.(\ref{eq:bree1}-\ref{eq:brpp1}) by roughly a factor of $2-5$.
It is not difficult to understand such situation: the large decrease is induced by the great
changes of the updated Gegenbauer moments in the wave functions.
 The updated Gegenbauer moment $a_2^\pi=0.115$ is indeed much smaller than the previous one
 $0.44$ for the leading twist-2 distribution amplitude, a factor of four decrease in magnitude.
For Gegenbauer moment $a_4^\pi$, which governs the high order contributions in pion DA,
it changes from $a_4^\pi =0.25$ to $a_4^\pi=-0.015$, even the sign is changed yet.
The form factor $F_{0,1}^{B \to \eta^{(\prime)}}$ then reduces from $0.30$ to $0.27$, which leads to a smaller
branching ratio. Besides the effect of a smaller form factor $F_{0,1}^{B \to \eta^{(\prime)}}$
which were extracted from the Fig.~1(a) and 1(b), the contributions to the branching ratios
from the remaining Feynman diagrams 1(c)-1(h) also become smaller than before due to the significant
variations of $a_2^\pi$ and $a_4^\prime$.
The total decrease of the branching ratios
therefore becomes significant. This fact tell us that current theoretical predictions still have
a strong dependence on the form of meson wave functions.

\begin{figure}[tb]
\centerline{\mbox{\epsfxsize=9cm\epsffile{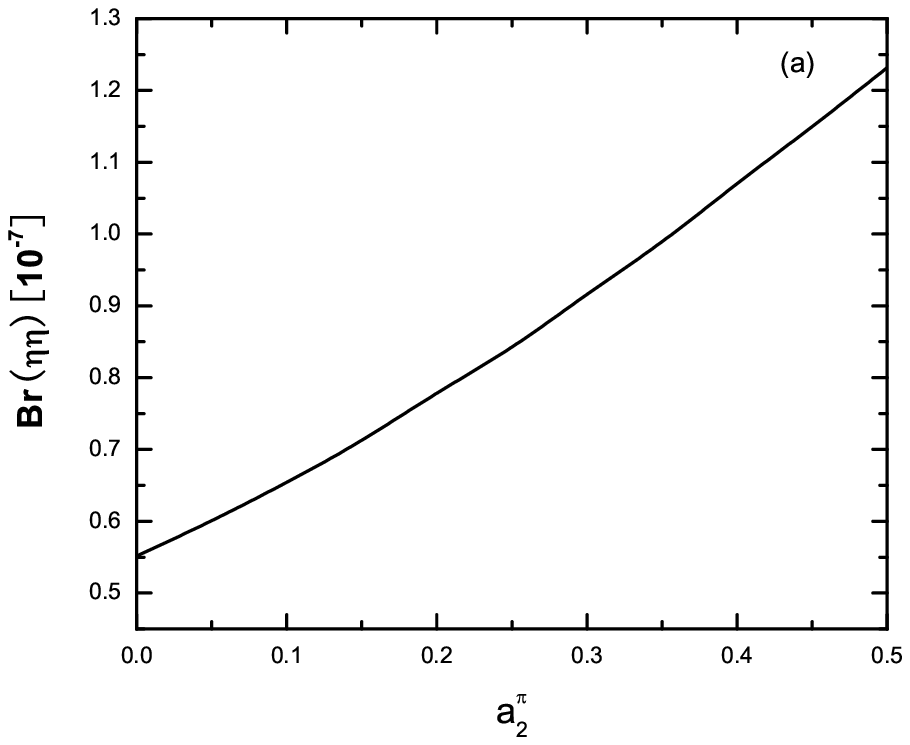}\epsfxsize=9cm\epsffile{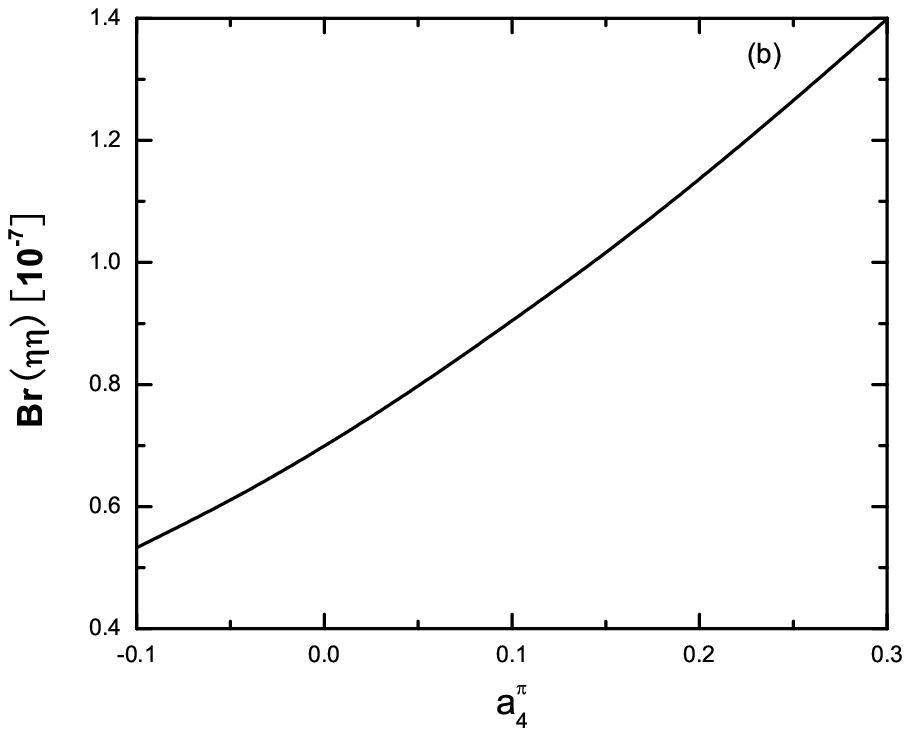}}}
\vspace{0.3cm}
 \caption{$a_2^\pi$- and $a_4^\pi$-dependence of the CP averaged branching ratio (in units of $10^{-7}$)
 $Br(B \to \eta \eta)$ for $0 \leq a_2^\pi \leq 0.50$ (a) and $-0.10 \leq a_4^\pi \leq 0.30$ (b).}
 \label{fig:fig5}
\end{figure}

We now take $B \to \eta \eta$ decay as an example to show the $a_2^\pi$- and $a_4^\pi$ -dependence
of the theoretical predictions explicitly.
Firstly, if we vary only the value of Gegenbauer moment $a_2^\pi$ in the range of
$0 \leq a_2^\pi \leq 0.50$, the pQCD prediction will be
\beq
0.55\times 10^{-7} \leq Br(B \to \eta \eta) \leq 1.23 \times 10^{-7}.
\eeq
This can be seen more clearly in Fig.~\ref{fig:fig5}a, where the CP averaged branching
ratio $Br(B \to \eta \eta)$ shows a linear dependence on $a_2^\pi$.
For the parameter $a_4^\pi$, we also find the similar linear dependence:
\beq
0.53\times 10^{-7} \leq Br(B \to \eta \eta) \leq 1.40 \times 10^{-7}
\eeq
for $-0.10 \leq a_4^\pi \leq 0.30$.

Although the pQCD predictions for the branching ratios obtained by employing the previous or updated
pion DAs are all consistent with the measured values due to still large theoretical and experimental errors,
the pQCD predictions in Eqs.(\ref{eq:bree3}-\ref{eq:brpp3}) are more reliable, in  our opinion,
since the updated Gegenbauer moments as given in Ref.~\cite{pball} are used to obtain these
results. Of course, better wave functions of particles involved are clearly needed to reduce the
errors of theoretical predictions.

In next section, we will use the updated Gengenbauer moments $(a_2^\pi,a_4^\pi)=(0.115,-0.015)$ to calculate
the CP-violating asymmetries for the three considered decay modes.

\subsection{CP-violating asymmetries }

Now we turn to the evaluations of the CP-violating asymmetries of $B \to \etap \etap$ decays in pQCD approach.
We here use the wave functions as presented in Appendix \ref{sec:app2}, but
with the updated Gengenbauer moments $(a_2^\pi,a_4^\pi)=(0.115,-0.015)$.
Because these decays are neutral B meson decays, so we should consider the effects of
$B^0-\bar{B}^0$ mixing. For $B^0$ meson decays into a CP eigenstate $f$,
the time-dependent CP-violating asymmetry can be defined as
\beq
\frac{Br \left (\overline{B}^0(t) \to f \right) - Br\left(B^0(t) \to f\right )}{
Br\left (\overline{B}^0(t) \to f\right ) + Br\left (B^0(t) \to f\right ) }
\equiv \acp^{dir} \cos
(\Delta m  \; t) + \acp^{mix} \sin (\Delta m \; t),
\label{eq:acp-def}
\eeq
where $\Delta m$ is the mass difference
between the two $B_d^0$ mass eigenstates, $ t =t_{CP}-t_{tag}
$ is the time difference between the tagged $B^0$ ($\overline{B}^0$)
and the accompanying $\overline{B}^0$ ($B^0$) with opposite b flavor
decaying to the final CP-eigenstate $f_{CP}$ at the time $t_{CP}$.
The direct and mixing induced CP-violating asymmetries
$\acp^{dir}$ and $\acp^{mix}$ can be written as
\beq
\acp^{dir}=\frac{ \left | \lambda_{CP}\right |^2 -1 }
{1+|\lambda_{CP}|^2}, \qquad A_{CP}^{mix}=\frac{ 2Im
(\lambda_{CP})}{1+|\lambda_{CP}|^2}, \label{eq:acp-dm}
\eeq
where the CP-violating parameter $\lambda_{CP}$ is
\beq
\lambda_{CP} =
\frac{ V_{tb}^*V_{td} \langle f |H_{eff}| \overline{B}^0\rangle} {
V_{tb}V_{td}^* \langle f |H_{eff}| B^0\rangle} = e^{2i\alpha}\frac{
1+z e^{i(\delta-\alpha)} }{ 1+ze^{i(\delta+\alpha)} }.
\label{eq:lambda2}
\eeq
Here the ratio $z$ and the strong phase
$\delta$ have been defined previously. In pQCD approach, since both
$z$ and $\delta$ are calculable, it is easy to find the numerical
values of $\acp^{dir}$ and $\acp^{mix}$ for the considered decay processes.

By using the mixing parameters in Eq.~(\ref{eq:t1-t8}) and the input parameters as given in Appendix
B, one found the pQCD predictions (in units of $10^{-2}$) for the direct and mixing
induced CP-violating asymmetries of the considered decays
\beq
\acp^{dir}(B^0 \to\eta \eta) &=& -33^{+2.6}_{-2.8}(\alpha)
^{+4.1}_{-3.8}(\omega_b) ^{+3.5} _{-0.0}(m_0^\pi),   \non
\acp^{mix}(B^0 \to\eta \eta) &=& +53.5^{+0.0}_{-3.4}(\alpha)
^{+3.1}_{-2.7}(\omega_b) ^{+2.1} _{-0.1}(m_0^\pi) ,  \\
\acp^{dir}(B^0 \to \eta \eta^\prime) &=& + 77.4^{+0.0}_{-5.6}
(\alpha) ^{+6.9}_{-11.2}(\omega_b)  ^{+8.0}_{-9.0}(m_0^\pi) ,  \non
\acp^{mix}(B^0\to \eta \eta^\prime) &=& -13.1^{+54.7}_{-48.8}
(\alpha) ^{+9.0}_{-9.9}(\omega_b) ^{+10.0}_{-6.2} (m_0^\pi),  \\
\acp^{dir}(B^0 \to \eta^\prime \eta^\prime) &=& +23.7
^{+10.0}_{-6.9} (\alpha) ^{+18.5}_{-16.9}(\omega_b)^{+6.0}_{-8.5}
(m_0^\pi),  \non
\acp^{mix}(B^0 \to \eta^\prime \eta^\prime) &=&
+93.2 ^{+4.9}_{-2.4} (\alpha) ^{+5.2}_{-11.1}(\omega_b)
^{+2.2}_{-2.1} (m_0^\pi),
\eeq
where the dominant errors come from the variations of $\alpha=100^\circ \pm
20^\circ$, $\omega_b=0.4\pm 0.04$ GeV and  $m_0^\pi=1.4\pm 0.1$ GeV.

\begin{figure}[tb]
\centerline{\mbox{\epsfxsize=9cm\epsffile{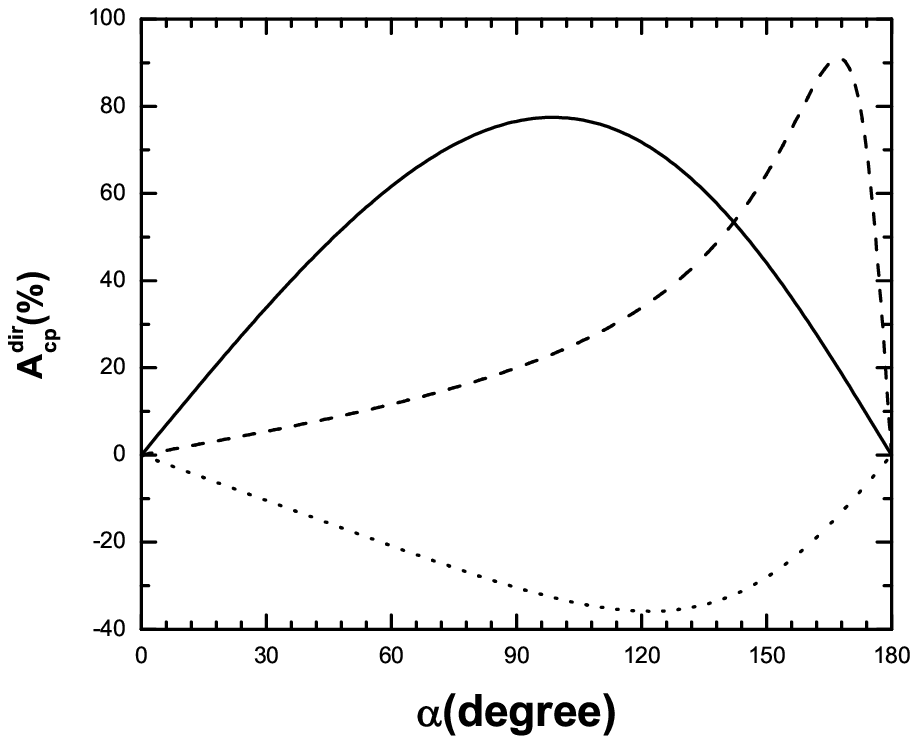}\epsfxsize=9cm\epsffile{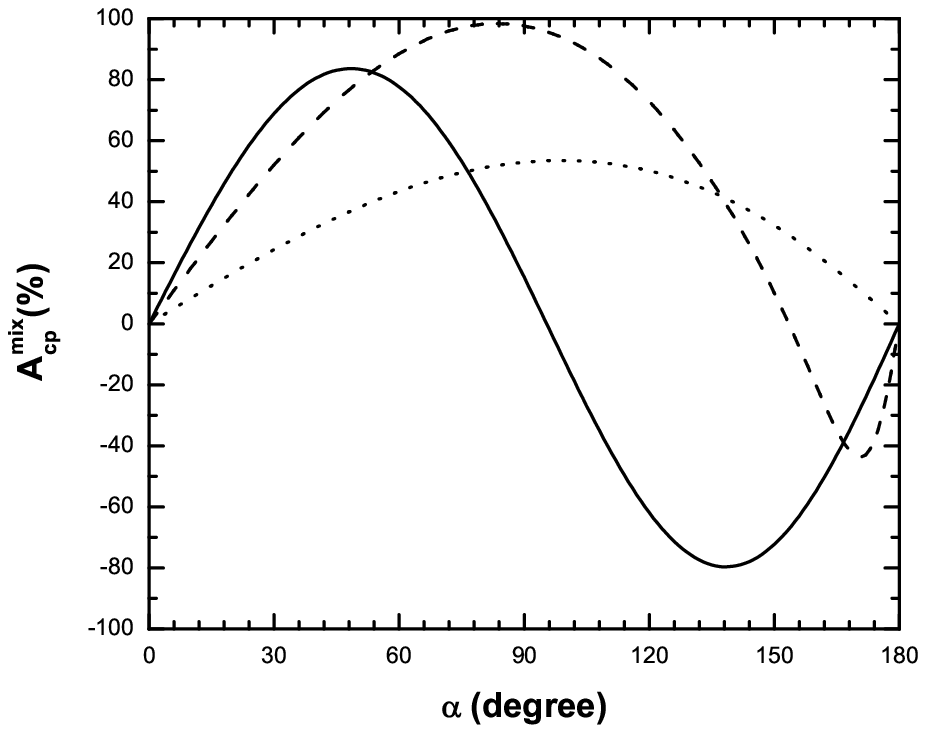}}}
\vspace{-0.5cm} \caption{The direct and mixing-induced CP asymmetry
(in percentage) of $B^0\to \eta \eta$ (dotted curve), $\eta
\eta^{\prime}$ (solid curve)  and $\eta^{\prime}
\eta^{\prime}$(dashed curve) decay as a function of CKM angle
$\alpha$.} \label{fig:fig6}
\end{figure}

In Fig.~\ref{fig:fig6}, we show the
$\alpha-$dependence of the pQCD predictions for the direct and the mixing-induced
CP-violating asymmetry for $B^0 \to \eta \eta$ (dotted curve), $B^0 \to  \eta  \eta^\prime$ (solid
curve) and $B^0 \to  \eta^\prime \eta^\prime$ (dashed curve) decay, respectively.

As a comparison, we present the theoretical predictions for $\acp^{dir}(B^0 \to \etap \etap)$ (in units of
$10^{-2}$ ) in both
the QCDF approach \cite{bn03b} \footnote{There is a sign difference between the term $\acp^{dir}$
defined here and the term $C_f$ defined in Ref.~\cite{bn03b}: i.e., $\acp^{dir}=-C_f$. }
and the SCET approach ~\cite{wz0601}
\beq
\acp^{dir}(B^0 \to \eta\eta) &=& \left\{ \begin{array}{ll}
+ 63{^{+32}_{-74}}, &\ \ \ { \rm QCDF}, \\
+ 48  \pm 32 , & \ \ \ {\rm SCET}, \\
\end{array} \right. \label{eq:acp11}\\
\acp^{dir}(B^0 \to \eta \eta^\prime) &=& \left\{ \begin{array}{ll}
+ 56{^{+32}_{-144}}, &\ \ \ { \rm QCDF}, \\
+ 70  \pm 24 , & \ \ \ {\rm SCET}, \\
\end{array} \right. \label{eq:acp12}\\
\acp^{dir}(B^0 \to \eta^\prime \eta^\prime) &=& \left\{ \begin{array}{ll}
+ 46{^{+43}_{-147}}, &\ \ \ { \rm QCDF}, \\
+ 60  \pm 38 , & \ \ \ {\rm SCET}, \\
\end{array} \right. \label{eq:acp24}
 \eeq
where the individual errors as given in Refs.~\cite{bn03b} and \cite{wz0601} have been added in quadrature.
From above numerical results one can see that
\begin{itemize}
\item[]{(i)}
In the considered three kinds of factorization approaches, the theoretical predictions for the
CP violating asymmetries are generally large in magnitude, and consistent with
each other if one takes the very large theoretical uncertainty into account.

\item[]{(ii)}
For $B \to \eta \eta$ decay, the pQCD prediction for $\acp^{dir}$ has an opposite
sign with those given in QCDF and SCET approach, which may be tested in the future B experiments.

\end{itemize}

If we integrate the time variable $t$, we will get the total CP
asymmetry for $B^0 \to \etap \etap$ decays,
\beq \acp
=\frac{1}{1+x^2} \acp^{dir} + \frac{x}{1+x^2} \acp^{mix} ,
\eeq
where $x=\Delta m/\Gamma=0.771$ for the $B^0-\overline{B}^0$ mixing
\cite{pdg04}.  Numerically, we found (in units of $10^{-2}$) that
\beq
 \acp^{tot}(B^0 \to \eta\eta)&=& 5.2 ^{+2.3}_{-3.4}(\alpha)
^{+4.0}_{-3.7}(\omega_b) ^{+3.2}_{-0.0} (m_0^\pi) ,  \\
\acp^{tot}(B^0 \to \eta \eta^\prime)&=&
42.2^{+24.0}_{-27.1}(\alpha)^{+8.7}_{-11.9}(\omega_b)^{+2.1}_{-0.8}(m_0^\pi), \\
\acp^{tot}(B^0 \to  \eta^\prime
\eta^\prime)&=&59.9^{+0.0}_{-3.5}(\alpha)^{+6.3}_{-8.0}(\omega_b)^{+2.8}_{-4.2}(m_0^\pi).
\eeq

\subsection{Effects of possible gluonic component of $\eta^\prime$}

Up to now, we have not considered the possible contributions to the
branching ratios and CP-violating asymmetries of $B \to \etap
\etap$ decays induced by the possible gluonic component of
$\eta^\prime$ \cite{ekou1,ekou2,rosner83}. When a non-zero
gluonic component exist in $\etapr$ meson, an additional decay amplitude ${\cal M}'$ will
be produced. Such decay amplitude may
interfere constructively  or destructively with the ones from the $q\bar{q}$ ($q=u,d,s$)
components of $\eta^\prime$, the branching ratios of the decays in
question may be increased or decreased accordingly.

In Ref.~\cite{bn03}, Beneke and Neubert computed the leading two-gluon contribution to the $B \to \etap$
form factors using the framework of QCD factorization.
In Ref.~\cite{ekou1}, Kou examined the gluonic component of $\eta^\prime$ and the contributions to
the process $gg\to \eta^\prime$. In his paper \cite{ekou1}, the $\eta$ and $\eta^\prime$ meson
were written as
\beq
|\eta> &=& X_\eta | \eta_q> + Y_\eta |\eta_s>, \non
|\eta^\prime> &=& X_{\eta^\prime} | \eta_q> + Y_{\eta^\prime} |\eta_s> + Z_{\eta^\prime} |gluonium>,
\eeq
where $\eta_q=(u\bar{u} + d\bar{d})/\sqrt{2}$ and $\eta_s=s\bar{s}$. From the experimental data on the
radiative light meson decays, such as $\phi \to \eta^\prime \gamma, \eta^\prime \to (\phi, \rho, \gamma) \gamma$
and $J/|psi \to \eta^\prime \gamma$ decays, the author found that
the gluonic component in $\eta^\prime$ should be less than $26\%$.

In Ref.~\cite{li0609}, by employing the pQCD factorization approach,  Charng, Kurimoto and Li
calculated the flavor-singlet contribution to the $B \to \etap$ transition form factors
induced by the Feynman diagrams with the two gluons emitted from the light
quark of the B meson (see Fig.~1 of Ref.~\cite{li0609}),
and they found that this gluonic contribution is negligible ($\sim 5\%$) in the $B \to \eta$
form factors, but can reach $10\% - 40\%$ in the $B \to \eta^\prime$ ones in the whole parameter
space. Such enhancement to $B \to \eta^\prime$ transition form factor can be help to explain the
large branching ratio $Br(B \to K \eta^\prime)$.

In order to check the gluonic effects on the decay modes under study,  we here follow the same procedure as
being used in Ref.~\cite{li0609} to include the possible gluonic contributions
to the $B \to \etap$ transition form
factors $F_{0,1}^{B\to \etap}$ and in turn to the branching ratios and CP violating asymmetries.
We found that the gluonic contributions to $B \to \etap \etap$ decays are rather small:
\begin{itemize}
\item
The central values of the pQCD predictions for the branching ratios
remain basically unchanged ( the variation is less than $2\%$ )
for $B \to \eta \eta$ and $B \to \eta^\prime \eta^\prime$
decay.  For $B \to \eta \eta^\prime$ decay, the enhancement is only about $5\%$:
the central value is changed from $0.60\times 10^{-7}$ to $0.63\times 10^{-8}$.

\item
As for the CP-violating asymmetries of $B \to \etap \etap $ decays, the possible gluonic corrections
are largely canceled in the ratio, and therefore negligible (less than $3\%$).

\item
The smallness of the gluonic corrections to the branching ratios can be understood as follows:
(a) the gluonic correction to $B \to \eta $ transition form factor itself is negligibly small \cite{li0609};
(b) only the first two diagram Fig.~1(a) and 1(b) are affected by the gluonic corrections to
$B \to \eta^\prime $ form factor, while the contributions from other six diagrams remain unchanged;
and (c) the total effects are consequently small.

\end{itemize}

Although much progress have been achieved in recent years, but frankly speaking, we currently still
do not know how to calculate reliably the contribution of the possible gluonic component
in $\eta^\prime$ meson.
From our previous works, as presented in Refs.~\cite{liu06,wang06} where only the dominant contributions
from quark contents of $\eta$ and $\eta^\prime$ were taken into account, the pQCD predictions
for the branching ratios of $B \to \rho \etap$ and $B \to \pi \etap$ decays also show a very good
agreement with currently available data. It seems that large gluonic contributions are unnecessary
for these decay modes. For $B \to K \eta^\prime$ decays, on the contrary, the gluonic
contribution may play an important role in explaining the so-called $K \eta^\prime$ puzzle \cite{bn03,li0609}.

Of course, more theoretical studies about the effects of possible gluonic component in $\eta^\prime$
and better experimental measurements for the relevant decay modes are needed to clarify this
point.

\section{summary }

In this paper,  we calculated the branching ratios and CP-violating
asymmetries of  $B^0 \to \eta \eta$, $ \eta \eta^\prime$ and $ \eta^{\prime}\eta^{\prime}$
decays at the leading order by using the pQCD factorization approach.
Besides the usual factorizable diagrams, the non-factorizable and
annihilation diagrams are also calculated analytically in the pQCD approach.
Furthermore, the annihilation diagrams  provide the
necessary strong phase required by a non-zero CP-violating  asymmetry for the considered decays.

From our calculations and phenomenological analysis, we found the following results:
\begin{itemize}
\item
The pQCD predictions for the form factors of $B \to \eta$ and $\eta^\prime$  transitions
agree well with those obtained in QCD sum rule calculations \cite{ball,pball}.

\item
Using the two mixing angle scheme and the updated Gegenbauer moments $a_2^\pi$ and $a_4^\pi$ ,
the pQCD predictions for the CP-averaged branching ratios are
\beq
Br(B^0 \to \eta\eta ) &=&  \left ( 0.67 ^{+0.32}_{-0.25} \right ) \times 10^{-7}, \\
Br(B^0 \to \eta\eta^{\prime}) &=&\left (0.18 \pm 0.11  \right ) \times 10^{-7}, \\
Br(B^0 \to\eta^{\prime}\eta^{\prime}  ) &=&\left (0.11 ^{+0.12}_{-0.09} \right ) \times 10^{-7},
\eeq
where the various errors as given in Eqs.~(\ref{eq:bree3}-\ref{eq:brpp3})
have been added in quadrature. The leading pQCD predictions are consistent with currently
available data, but both the theoretical and experimental errors are still large.

\item
For the CP-violating asymmetries of the considered three decay modes, the pQCD predictions
are generally large in magnitude, and have large theoretical uncertainty.

\end{itemize}

\begin{acknowledgments}

We are very grateful to Cai-Dian L\"u, Ying Li, Xin Liu and Huisheng Wang for helpful discussions.
This work is partly supported  by the National Natural Science Foundation of China under Grant
No.10275035,10575052, and by the Specialized Research Fund for the doctoral Program of higher
education (SRFDP) under Grant No.~20050319008.

\end{acknowledgments}

%%%%%%%%%%%%%%%%%%%%%%%%%%%%%%%%%%%%%%%%%%%%%%%%%%%%%%%%%%%%%%%%%%%%%%%%%%%%%%%%%%
%                                        Appendix
%%%%%%%%%%%%%%%%%%%%%%%%%%%%%%%%%%%%%%%%%%%%%%%%%%%%%%%%%%%%%%%%%%%%%%%%%%%%%%%%5

\begin{appendix}

\section{Related Functions }\label{sec:app1}

We show here the function $h_i$'s, coming from the Fourier
transformations  of the function $H^{(0)}$,
\beq
 h_e(x_1,x_3,b_1,b_3)&=&
 K_{0}\left(\sqrt{x_1 x_3} m_B b_1\right)
 \left[\theta(b_1-b_3)K_0\left(\sqrt{x_3} m_B
b_1\right)I_0\left(\sqrt{x_3} m_B b_3\right)\right.
 \non
& &\;\left. +\theta(b_3-b_1)K_0\left(\sqrt{x_3}  m_B b_3\right)
I_0\left(\sqrt{x_3}  m_B b_1\right)\right] S_t(x_3), \label{he1}
\eeq
 \beq
 h_a(x_2,x_3,b_2,b_3)&=&
 K_{0}\left(i \sqrt{x_2 x_3} m_B b_2\right)
 \left[\theta(b_3-b_2)K_0\left(i \sqrt{x_3} m_B
b_3\right)I_0\left(i \sqrt{x_3} m_B b_2\right)\right.
 \non
& &\;\;\;\;\left. +\theta(b_2-b_3)K_0\left(i \sqrt{x_3}  m_B
b_2\right) I_0\left(i \sqrt{x_3}  m_B b_3\right)\right] S_t(x_3),
\label{he3} \eeq
 \beq
 h_{f}(x_1,x_2,x_3,b_1,b_2) &=&
 \biggl\{\theta(b_2-b_1) \mathrm{I}_0(M_B\sqrt{x_1 x_3} b_1)
 \mathrm{K}_0(M_B\sqrt{x_1 x_3} b_2)
 \non
&+ & (b_1 \leftrightarrow b_2) \biggr\}  \cdot\left(
\begin{matrix}
 \mathrm{K}_0(M_B F_{(1)} b_2), & \text{for}\quad F^2_{(1)}>0 \\
 \frac{\pi i}{2} \mathrm{H}_0^{(1)}(M_B\sqrt{|F^2_{(1)}|}\ b_2), &
 \text{for}\quad F^2_{(1)}<0
\end{matrix}\right),
\label{eq:pp1}
 \eeq
\beq
h_f^3(x_1,x_2,x_3,b_1,b_2) &=& \biggl\{\theta(b_1-b_2)
\mathrm{K}_0(i \sqrt{x_2 x_3} b_1 M_B)
 \mathrm{I}_0(i \sqrt{x_2 x_3} b_2 M_B)+(b_1 \leftrightarrow b_2) \biggr\}
 \non
& & \cdot
 \mathrm{K}_0(\sqrt{x_1+x_2+x_3-x_1 x_3-x_2 x_3}\ b_1 M_B),
 \label{eq:pp4}
\eeq
 \beq
 h_f^4(x_1,x_2,x_3,b_1,b_2) &=&
 \biggl\{\theta(b_1-b_2) \mathrm{K}_0(i \sqrt{x_2 x_3} b_1 M_B)
 \mathrm{I}_0(i \sqrt{x_2 x_3} b_2 M_B)
 \non
&+& (b_1 \leftrightarrow b_2) \biggr\} \cdot \left(
\begin{matrix}
 \mathrm{K}_0(M_B F_{(2)} b_1), & \text{for}\quad F^2_{(2)}>0 \\
 \frac{\pi i}{2} \mathrm{H}_0^{(1)}(M_B\sqrt{|F^2_{(2)}|}\ b_1), &
 \text{for}\quad F^2_{(2)}<0
\end{matrix}\right), \label{eq:pp3}
\eeq
where $J_0$ is the Bessel function and  $K_0$, $I_0$ are
modified Bessel functions $K_0 (-i x) = -(\pi/2) Y_0 (x) + i
(\pi/2) J_0 (x)$, and $F_{(j)}$'s are defined by
\beq
F^2_{(1)}&=&(x_1 -x_2) x_3\;,\\
F^2_{(2)}&=&(x_1-x_2) x_3\;\;.
 \eeq

The threshold resummation form factor $S_t(x_i)$ is adopted from
Ref.\cite{kurimoto}
\beq S_t(x)=\frac{2^{1+2c} \Gamma
(3/2+c)}{\sqrt{\pi} \Gamma(1+c)}[x(1-x)]^c,
\eeq
where the parameter $c=0.3$. This function is normalized to unity.

The Sudakov factors used in the text are defined as
\beq
S_{ab}(t) &=& s\left(x_1 m_B/\sqrt{2}, b_1\right) +s\left(x_3 m_B/\sqrt{2},
b_3\right) +s\left((1-x_3) m_B/\sqrt{2}, b_3\right) \non
&&-\frac{1}{\beta_1}\left[\ln\frac{\ln(t/\Lambda)}{-\ln(b_1\Lambda)}
+\ln\frac{\ln(t/\Lambda)}{-\ln(b_3\Lambda)}\right],
\label{wp}\\
S_{cd}(t) &=& s\left(x_1 m_B/\sqrt{2}, b_1\right)
 +s\left(x_2 m_B/\sqrt{2}, b_2\right)
+s\left((1-x_2) m_B/\sqrt{2}, b_2\right) \non
 && +s\left(x_3
m_B/\sqrt{2}, b_1\right) +s\left((1-x_3) m_B/\sqrt{2}, b_1\right)
\non
 & &-\frac{1}{\beta_1}\left[2
\ln\frac{\ln(t/\Lambda)}{-\ln(b_1\Lambda)}
+\ln\frac{\ln(t/\Lambda)}{-\ln(b_2\Lambda)}\right],
\label{Sc}\\
S_{ef}(t) &=& s\left(x_1 m_B/\sqrt{2}, b_1\right)
 +s\left(x_2 m_B/\sqrt{2}, b_2\right)
+s\left((1-x_2) m_B/\sqrt{2}, b_2\right) \non
 && +s\left(x_3
m_B/\sqrt{2}, b_2\right) +s\left((1-x_3) m_B/\sqrt{2}, b_2\right)
\non
 &
&-\frac{1}{\beta_1}\left[\ln\frac{\ln(t/\Lambda)}{-\ln(b_1\Lambda)}
+2\ln\frac{\ln(t/\Lambda)}{-\ln(b_2\Lambda)}\right],
\label{Se}\\
S_{gh}(t) &=& s\left(x_2 m_B/\sqrt{2}, b_2\right)
 +s\left(x_3 m_B/\sqrt{2}, b_3\right)
+s\left((1-x_2) m_B/\sqrt{2}, b_2\right) \non
 &+& s\left((1-x_3)
m_B/\sqrt{2}, b_3\right)
-\frac{1}{\beta_1}\left[\ln\frac{\ln(t/\Lambda)}{-\ln(b_2\Lambda)}
+\ln\frac{\ln(t/\Lambda)}{-\ln(b_3\Lambda)}\right], \label{ww}
\eeq
where the function $s(q,b)$ are defined in the Appendix A of
Ref.\cite{luy01}. The scale $t_i$'s in the above equations are
chosen as
\beq
t_{e}^1 &=& {\rm max}(\sqrt{x_3} m_B,1/b_1,1/b_3)\;,\non
t_{e}^2 &=& {\rm max}(\sqrt{x_1}m_B,1/b_1,1/b_3)\;,\non
t_{e}^3 &=& {\rm max}(\sqrt{x_3}m_B,1/b_2,1/b_3)\;,\non
t_{e}^4 &=& {\rm max}(\sqrt{x_2}m_B,1/b_2,1/b_3)\;,\non
t_{f} &=& {\rm max}(\sqrt{x_1 x_3}m_B, \sqrt{(x_1-x_2) x_3}m_B,1/b_1,1/b_2)\;,\non
t_{f}^3 &=& {\rm max}(\sqrt{x_1+x_2+x_3-x_1 x_3-x_2 x_3}m_B,
    \sqrt{x_2 x_3} m_B,1/b_1,1/b_2)\;,\non
t_{f}^4 &=&{\rm max}(\sqrt{x_2 x_3} m_B,\sqrt{(x_1-x_2)
x_3}m_B,1/b_1,1/b_2)\; .
\eeq
They are chosen as the maximum energy scale appearing in each diagram
to kill the large logarithmic radiative corrections.

\section{Input parameters and wave functions} \label{sec:app2}

In this Appendix we show the input parameters and the light meson wave functions to be used in the
numerical calculations.

The masses, decay constants, QCD scale  and $B^0$ meson lifetime are
\beq
 \Lambda_{\overline{\mathrm{MS}}}^{(f=4)} &=& 250 {\rm MeV}, \quad
 f_\pi = 130 {\rm MeV}, \quad  f_B = 190 {\rm MeV}, \non
 m_0^{\eta_{d\bar{d}}}&=& 1.4 {\rm GeV},\quad m_{0}^{\eta_{s\bar{s}}}=2.4 {\rm
 GeV}, \quad m_\pi =140 {\rm MeV}, \quad
  f_K = 160  {\rm MeV}, \non
 M_B &=& 5.2792 {\rm GeV}, \quad M_W = 80.41{\rm
 GeV},\tau_{B^{0}}=1.54\times10^{-12}{\rm s}
 \label{para}
\eeq

The central values of the CKM matrix elements as given in Ref.~\cite{pdg04} are
\beq
|V_{ud}|&=&0.9745, \quad |V_{ub}|=0.0040,\non
|V_{tb}|&=&0.9990, \quad |V_{td}|=0.0075.
\eeq

For the $B$ meson wave function, we adopt the model
\beq
\phi_B(x,b) &=& N_B x^2(1-x)^2 \mathrm{exp} \left
 [ -\frac{M_B^2\ x^2}{2 \omega_{b}^2} -\frac{1}{2} (\omega_{b} b)^2\right],
 \label{phib}
\eeq
where $\omega_{b}$ is a free parameter and we take
$\omega_{b}=0.4\pm 0.04$ GeV in numerical calculations, and
$N_B=91.745$ is the normalization factor for $\omega_{b}=0.4$. This
is the same wave functions as being used in Refs.~\cite{luy01,kls01,kurimoto},
which is a best fit for most of the measured hadronic B decays.

For the distribution amplitudes $\phi_{\eta_{d\bar{d}}}^A$,
$\phi_{\eta_{d\bar{d}}}^P$ and $\phi_{\eta_{d\bar{d}}}^T$ appeared in Eq.~(\ref{eq:ddbar}),
we utilize the result from the light-cone sum rule~\cite{ball,bf} including twist-3 contribution:
\beq
\phi_{\eta_{d\bar{d}}}^A(x)&=&\frac{3}{\sqrt{2N_c}}f_xx(1-x)
\left\{ 1+a_2^{\eta_{d\bar{d}}}\frac{3}{2}\left [5(1-2x)^2-1
\right ]\right. \non &&\left. + a_4^{\eta_{d\bar{d}}}\frac{15}{8}
\left [21(1-2x)^4-14(1-2x)^2+1 \right ]\right \},  \non
\phi^P_{\eta_{d\bar{d}}}(x)&=&\frac{1}{2\sqrt{2N_c}}f_x \left \{
1+ \frac{1}{2}\left (30\eta_3-\frac{5}{2}\rho^2_{\eta_{d\bar{d}}}
\right ) \left [ 3(1-2x)^2-1 \right] \right.  \non
&& \left. + \frac{1}{8}\left
(-3\eta_3\omega_3-\frac{27}{20}\rho^2_{\eta_{d\bar{d}}}-
\frac{81}{10}\rho^2_{\eta_{d\bar{d}}}a_2^{\eta_{d\bar{d}}} \right
) \left [ 35 (1-2x)^4-30(1-2x)^2+3 \right ] \right\} ,  \non
\phi^T_{\eta_{d\bar{d}}}(x) &=&\frac{3}{\sqrt{2N_c}}f_x(1-2x) \non
 && \cdot \left [ \frac{1}{6}+(5\eta_3-\frac{1}{2}\eta_3\omega_3-
\frac{7}{20}\rho_{\eta_{d\bar{d}}}^2
-\frac{3}{5}\rho^2_{\eta_{d\bar{d}}}a_2^{\eta_{d\bar{d}}})(10x^2-10x+1)\right
],  \non
\eeq
with \cite{ball,kf}
\beq
a^{\eta_{d\bar{d}}}_2&=& a_2^\pi=  0.44, \quad a^{\eta_{d\bar{d}}}_4=a_4^\pi = 0.25,\non
\rho_{\eta_{d\bar{d}}}&=&m_{\pi}/{m_0^{\eta_{d\bar{d}}}}, \quad
\eta_3=0.015, \quad \omega_3=-3.0.\label{eq:a2pi}
\eeq
The updated Gegenbauer moments as given in Ref.~\cite{pball} are
\beq
a_2^\pi= 0.115, \qquad a_4^\pi=-0.015.
\eeq

We also assume that the wave function of $u\bar{u}$ is the same as the wave
function of $d\bar{d}$ \cite{ekou2}.
For the wave function of the $s\bar{s}$ components, we also use the
same form as $d\bar{d}$ but with $m^{s\bar{s}}_0$ and $f_y$ instead
of $m^{d\bar{d}}_0$ and $f_x$, respectively. For $f_x$ and $f_y$, we
use the values as given in Ref.~\cite{kf} where isospin symmetry
is assumed for $f_x$ and $SU(3)$ breaking effect is included for
$f_y$:
 \beq
 f_x=f_{\pi}, \ \ \ f_y=\sqrt{2f_K^2-f_{\pi}^2}.\ \ \
\label{eq:fxfy}
\eeq

These values are translated to the values in the two mixing angle
scheme, which is often used in vacuum saturation approach as:
\beq
f_1&=& 151  {\rm MeV}, \quad f_8 =169  {\rm MeV}, \non
\theta_1 &= & -2.4^\circ, \quad \theta_8= -21.2^{\circ}. \quad
\eeq
The parameters $m_0^i$ $(i=\eta_{d\bar{d}(u\bar{u})}, \eta_{s\bar{s}})$
are defined as:
\beq
m_0^{\eta_{d\bar{d}(u\bar{u})}}\equiv m_0^\pi
\equiv \frac{m_{\pi}^2}{(m_u+m_d)}, \qquad
m_0^{\eta_{s\bar{s}}}\equiv \frac{2M_K^2-m_{\pi}^2}{(2m_s)}.
\eeq

\end{appendix}

%%%%%%%%%%%%%%%%%%%%%%%%%%%%%%%%%%%%%%%%%%%%%%%%%%%%%%%%%%%%%%%%%%%%%%%%%%%%%%%%%%%%%%%%%%%%%%5
%                                 reference
%%%%%%%%%%%%%%%%%%%%%%%%%%%%%%%%%%%%%%%%%%%%%%%%%%%%%%%%%%%%%%%%%%%%%%%%%%%%%%%%%%%%%%%%%%%%%%%%%

\end{document}